\documentclass[iop,twocolappendix,apj,floatfix]{emulateapj}

  \usepackage{graphicx}                       % to insert files .ps and .eps
  \usepackage{amsfonts,amsmath,amssymb}
  \usepackage{bm}                             %.. bold math
  \usepackage[usenames,dvipsnames]{xcolor}    %.. OliveGreen
  \usepackage{natbib,color}
   \usepackage{enumitem}
    \newcommand{\changed}   [1] {{\textcolor{blue}{\textbf{{#1}}}}}
  \renewcommand{\changed}   [1] {#1}

    \newcommand{\remarkHide}[1]{{\textcolor{blue}{{\ldots \sf #1}}}\newline}
  \renewcommand{\remarkHide}[1]{}

  \renewcommand{\d} {\mathrm{d}}                 %..    Use in derivatives
    \newcommand{\PD}    [2]    {\frac{\partial{#1}} {\partial{#2}} }
    \newcommand{\avg}   [1]    {\langle #1 \rangle}

   \newcommand{\vct}[1]  {\ensuremath{\boldsymbol{#1}}}  %..     bold italic
 \newcommand{\zhat} {\hat{\vct{z}}}
 \newcommand{\ehat} {\hat{\vct{e}}}

 \newcommand{\vB} {\vct{B}}
 \newcommand{\vV} {\vct{V}}

 \newcommand{\va} {\vct{a}}
 \newcommand{\vb} {\vct{b}}
 \newcommand{\vj} {\vct{j}}
 \newcommand{\vv} {\vct{v}}
 \newcommand{\vx} {\vct{x}}
 \newcommand{\vz} {\vct{z}}
 \newcommand{\vk} {\vct{k}}

 \def\vr{\vct{r}}               %..     redefines without error(revtex)
   \newcommand{\pbeta} {\beta_\text{p}}

   \newcommand{\tauNL} {\tau_\text{nl}}
   \newcommand{\tauA}  {\tau_\text{A}}
   \newcommand{\tauS}  {\tau_\text{s}}
   \newcommand{\tauW}  {\tau_\text{wave}}

   \newcommand{\Higdelta} {\varepsilon}     %.. what Higdon84 calls delta

   \newcommand{\Emod}  {E^\text{3D}}
   \newcommand{\Eomni} {E^\text{omni}}
   \newcommand{\Epar}  {E^\parallel}
   \newcommand{\Ered}  {E^\text{red}}

   \newcommand{\ldiss} {\ell_\text{diss}}
   \newcommand{\kdiss} {k_\text{diss}}

\begin{document}

   \shorttitle{} %% \qquad \textsf{\color{blue}\today}}
   \shortauthors{}%% \qquad \textsf{\color{blue}\today}}     %..apj
%   \shortauthor{ \qquad \textsf{\color{blue}\today}}      %..jpp

  \title{Critical Balance and the Physics of MHD Turbulence}

\author{      S. Oughton$^1 $
        and
        W. H. Matthaeus$^2 $}

%% \affiliation{$^1 $%.. jpp
 \affil{$^1 $%.. apj
         Department of Mathematics and Statistics,
         University of Waikato, Hamilton 3240, New Zealand\\
        $^2 $%
         Bartol Research Institute,
         Department of Physics and Astronomy,
         University of Delaware, Delaware 19716, USA\\
}
%%\email{$^{\dagger}$whm@udel.edu}

        \date{}

%% \maketitle          %..jpp

\begin{abstract}
A discussion of the advantages and limitations of the concept of
critical balance,
as employed in turbulence phenomenologies, is
presented.  The incompressible magnetohydrodynamic (MHD) case is a
particular focus.  The discussion emphasizes the
status of the original
        \cite{GoldreichSridhar95}
critical balance conjecture
relative to related theoretical issues and models in
an MHD description of plasma turbulence.
Issues examined include variance and spectral anisotropy,
influence of a mean magnetic field, local and nonlocal effects,
and the potential for effects of external driving.
Related models such as Reduced MHD
provide a valuable context in the considerations.
Some new results concerning spectral features and timescales are
presented in the course of the discussion.
Also mentioned briefly are some
adaptations and variations of critical balance.
\end{abstract}
%====================================================================

  \maketitle            %.. apj

%... preface: the introductory remarks of a speaker/author

        \section{Introduction}
                \label{sec:preface}

A well-known sequence of papers
   \citep[referred to hereafter as SG94, GS95, and GS97]
        {SridharGoldreich94,GoldreichSridhar95,GoldreichSridhar97}
defined a theoretical approach to magnetohydrodynamic (MHD)
turbulence that has become known as
        \emph{critical balance}  (CB)
theory.
This approach has been widely cited in the astrophysics community.
   %%      \citep{REFS} \ldots
It has also gained considerable attention in the space plasma
community and has been suggested to be of general relevance for
turbulent systems with waves,
   %%      \citep[e.g.,][]{NazarenkoSchekochihin11},
including gyrokinetics.
   %%      \citep{SchekochihinEA09}.
Due in part to its wide adoption in applications,
CB is often viewed as a starting point for incompressible MHD
turbulence theory rather than a model that emerges from a subtle
series of approximations and physical arguments.
As a consequence, on the one hand,
the underpinnings of CB are not always subject to appropriate scrutiny
in determining its applicability,
while on the other hand, quantitative signatures are sometimes
attributed to CB when a less restrictive explanation is available.
What is apparently needed is a
review of the physical origins
of CB including a compendium of connections and relationships
to other models, such as reduced MHD,
a distinct model which itself ensues from a
nuanced sequence of considerations.
   %%   \citep[see, e.g.,][]{ZankMatt92-rmhd,OughtonEA17-rmhd}.
The utility of such a collection, as well as the potential to
stimulate more vigorous discussion of these topics,
provide ample motivation for the present
review.
%====================================================================

%   \vfill
%   \vspace*{17mm}
%   \tableofcontents
% \clearpage

%----------------------------------------------

        \section{Preliminaries}
                \label{sec:prelims}

MHD turbulence is of considerable relevance to
systems such as the solar corona, solar wind, planetary
magnetospheres, the interstellar medium, accretion discs, and
laboratory plasmas
        \citep{Parker-cmf,Parker-conversations,Moffatt,
                Barnes79a,MacLow99,
                BalbusHawley98,RobinsonRusbridge71,
                BrandenburgNordlund11,BrandenburgLazarian13,
                BeresnyakLazarian-mhdturb}.
Often a large-scale magnetic field is also present and this can have
significant impact on the turbulent dynamics
        \citep{Iroshnikov64,Kraichnan65}.
As analytic solutions for turbulent flows are
 in short supply,
%% difficult to obtain,
it is typical to employ instead
modelling of various kinds or numerical approximations.
Our interest here is in discussion of
        \emph{critical balance} (CB)
phenomenologies
        (GS95),
%%            \citep[hereafter GS95] {GoldreichSridhar95}.
%%                  NazarenkoSchekochihin11},
primarily as they are applied to a class of MHD
        \emph{inertial range} (IR) models.
A central tenet of CB
is that the IR dynamics is dominated by those wavevector
        ($ \vk $)
modes for which there is approximate equality of the wave timescale
        $ \tauW( \vk)$
and the nonlinear timescale
        $ \tauNL( |\vk| ) $;
that is,
%----------------------------------------------------------
\begin{equation}
         \tauW( \vk)  \approx  \tauNL( |\vk| )
    \label{eq:CB-condit}
\end{equation}
%----------------------------------------------------------
for the dynamically important wavevectors $\vk$.
This equation serves to define that region of wavevector space for
which this equality holds.  We refer to this as the
        \emph{equal timescale region}.
\changed{As is well-known,
        $ \tauNL $
        plays a crucial role in the spectral transfer (or cascade) of
        energy,
        and when CB is invoked so too does the wave timescale.}

Equation~(\ref{eq:CB-condit})
is often referred to as the
        \emph{critical balance condition}
and defines the
        \emph{equal timescale curve} (or zone).
It implies a relationship between the
        parallel
and     perpendicular
wavenumbers, $ k_z$ and $k_\perp$,
and typically this is anisotropic.
In the IR, the nonlinear and wave timescales are obviously
scale (or $\vk$)
dependent.
Still, if the IR is critically balanced, then at each scale
these are of similar magnitude:
        $ \tauNL( k)  \approx  \tauW( \vk) $.
It follows that the triple correlation and cascade timescales are also
of this magnitude
        \citep[e.g.,][]{Kraichnan65,ZhouEA04}.
Hence, from a quantitative standpoint, in homogeneous turbulence,
there is only one
        ($\vk$-dependent)
timescale of relevance to IR dynamics,
expressible equally well as either
        $ \tauNL $ or $ \tauW $.
It is this equivalence of timescales that simplifies development of
turbulence phenomenologies
when critical balance is invoked.

However, from the time of the earliest
statement of critical balance
        (GS95)
and its emergence from weak turbulence
        (SG94),
%%        \citep{SridharGoldreich94},
questions have been raised
concerning
its assumptions
and range of validity
  \citep{MontMatt95,NgBhattacharjee96,NgBhattacharjee97},
as well as its relationship to
Kolmogorov theory and
other models such as    Reduced MHD (RMHD).
These issues relate directly to the
subtleties and ambiguities in the application of critical balance as a
principle,
as we will discuss.

Indeed, our aim herein is to
present a detailed critique of the advantages and
limitations of CB phenomenologies, particularly as they are applied
to the case of incompressible MHD.
This will include discussion of what it means for turbulence to be strong.

The original context for CB,
as expounded by         GS95,
was incompressible MHD with a uniform mean magnetic
field,\footnote{Magnetic fields
               are measured in Alfv\'en velocity units, e.g.,
                 $ \vb \to \vb / \sqrt{4\pi \rho}$
               and $\vV_A \equiv \vB_0$.}
  $ \vB_0 $,
of \emph{moderate} strength.
%% with an energy density comparable to that of the fluctuations.
In this case, the wave timescale used in the CB condition is that for
(shear) Alfv\'en waves:
        $ \tauA ( \vk) = 1 / | \vk \cdot \vB_0| $.
Note the anisotropic dependence on   $ \vk $.
Subsequently,
CB approaches have been employed for MHD with various combinations of
stronger $ B_0 $,
anisotropy in the perpendicular plane,
and
nonzero cross helicity
        \citep[e.g.,][]{GaltierEA05,
                Boldyrev05,Boldyrev06,
                ChoEA02,
                LithwickGoldreich01,LithwickGoldreich03,LithwickEA07,
                Chandran08-Hc,
                BeresnyakLazarian08-Hc,BeresnyakLazarian09-Hc,
                PerezBoldyrev09,PerezBoldyrev10-apj,PerezBoldyrev10-pop,
                PodestaBhattacharjee10},
and other systems where nonlinear effects and linear waves might both
be active
%% , such as rotating or stratified hydrodynamics
        \citep[e.g.,][]{SchekochihinEA09,ChoLazarian09,
                        NazarenkoSchekochihin11,TenBargeHowes12,
                        Terry18}.

Much of the notation that we employ is summarized in
  Table~\ref{tab:notation}.
Throughout the paper,
perpendicular       %% ($ \perp $)
and parallel                    %% ($ \parallel $)
are relative to the direction
of the uniform mean magnetic field,     $ \vB_0 $,
as in GS95, unless otherwise noted.
(Local mean fields are discussed in \S\ref{sec:local-field}.)
The role of wave, or even wave-like, activity in turbulence is still
enjoying considerable debate.
Herein, we attempt to avoid biasing thinking towards wave
interpretations by preferring terms that do not presuppose the
presence of wave activity.
For example,
when discussing the polarizations of general incompressible fluctuations
we describe them as toroidal and poloidal
rather than Alfv\'enic and pseudo-Alfv\'enic,
since the latter pair suggest, strongly, that the fluctuations are
small amplitude waves.
The glossary defines these and related terms.

Throughout the discussion herein, except where noted, there is
an assumption that the (nonlinear) dynamical effects of interest are
governed by interactions that are \emph{local} in scale.
Accordingly
a simple estimate for the nonlinear timescale
is
        $ \tauNL( | \vk | ) \approx 1 / [ k \,\delta v(k)] $
where
        $ \delta v(k) $
is the average velocity fluctuation at scales
        $ \ell  \sim   1 / |\vk| $.
Full definitions of the timescales are given in later sections.

Our discussions of cascades and spectral transfer
will be mainly from a phenomenological perspective,
  \changed{and will typically
     assume that the turbulence is fully developed.
     This requires that the Reynolds numbers are large enough to
     support such states
        \citep[e.g.,][]{Zhou07,ZhouOughton11},
     as is often the case in astrophysical environments
     because of various instabilities,
     or forcing mechanisms like supernovae explosions
        \cite[e.g.,][]{BalbusHawley98,
                       Zhou17-i,Zhou17-ii,ZhouEA19-mixing,
                       BrandenburgNordlund11,BeresnyakLazarian-mhdturb}.
    For recent quantitative discussion of the asymptotic saturation of
    cascade rates at large Reynolds numbers in three dimensional (3D)
    MHD see \cite{LinkmannEA17-Ceps,BandyopadhyayEA18-Ceps}.
}
For more formal treatments of spectral transfer see references such as
        \cite{Zhou93-xfer,
                Alexakis07,AlexakisEA07-pre,AlexakisEA07-njp,
                Verma04,Verma19,
                DomaradzkiEA10,AluieEyink10,
                McComb-HIT}.
%%   ZhouOughton11

\changed{It is worth noting that there are many astrophysical systems
  with multiple
  distinct timescales that are \emph{not} necessarily comparable
  in magnitude, and for these critical balance approaches may not be
  very relevant.
  For example, turbulent mixing can occur,
  via fluid instabilities,
  at interfaces associated with magnetosphere--stellar wind
  boundaries and at supernovae shock fronts
     \cite[e.g.,][]{Zhou17-i,Zhou17-ii,ZhouEA19-mixing}.
  Other examples include turbulence in
  molecular clouds,
  the interstellar medium,
  and
  magnetic reconnection regions
         \citep[e.g.,][]{BruntEA09,
                ElmegreenScalo04,Higdon84,FraternaleEA19,
                LazarianEA12}
 }

In the following sections
we begin by discussing
antecedents of CB,
and delving into the
definitions of weak and strong turbulence.
We then examine, in   \S\ref{sec:emergence},
CB as presented and developed for the two situations considered in
GS95: the weak turbulence to strong turbulence case and the initially
strong turbulence case.
For each of these we start with a precis of the GS95
approach and then present relevant commentary and critique.
  Section~\ref{sec:local-field}
considers issues related to use of local mean fields rather than the
global mean field,
while   \S\ref{sec:rmhd}
compares the CB approach to the
relationships embodied in derivations of RMHD.
This leads to discussion of the wider applicability of the equal
timescale curve         (\S\ref{sec:curve})
and some of its properties      (\S\ref{sec:attract}).
   %% including a new perspective on spectral transfer in strong MHD
   %% turbulence.
Section~\ref{sec:corrn-length}
considers connections between (quasi-)2D fluctuations, parallel
correlation lengths, and the postulates of CB.
The relevance of non-toroidal fluctuations, discarded in GS95, is
discussed in    \S\ref{sec:otherFlucts}.
Observational and simulation support for CB is
considered in           \S\ref{sec:observations},
followed by a summary section.
Several appendices and a glossary close the paper.

%------------------------------------------------------------------------------%
  \begin{table} [tbp]
    \caption{Some notation. Where only velocity fluctuations are shown,
      analogous definitions also hold for the magnetic field fluctuations.}
        \label{tab:notation}
    \begin{center}
    \begin{tabular} {| c | l |} \hline
        $ \epsilon $    &  energy cascade rate (in inertial range)\\
        $ \vB_0 $       &  large-scale magnetic field (often uniform)\\
        $ k_z, \vk_\perp $
                        &  components of wavevector $\vk$ wrt global $\vB_0$ \\
        $ \vv, \vb $    &  velocity and magnetic field fluctuations\\
        $ \delta v $
                        &  (global) rms fluctuation strength for
                        $\vv$\\
        $ \delta v(\vk) $
                        & rms fluctuation in $\vv$ at scales
                        $\sim \ell \equiv 1/|\vk| $ \\
        $ \delta v_\perp(\vk) $
                        & rms fluctuation in $\vv_\perp$ at scale
                        $\sim 1/k $ \\
        $ \tauNL( \vk)$ &  nonlinear timescale associated with, e.g.,
                                $\vv\cdot \nabla \vv$\\
        $ \tauW( \vk) $ &  linear wave timescale (often anisotropic wrt $\vk$)\\
        $ \tauA( \vk) $   %%= 1/ | \vk \cdot \vB_0 |$
                        & Alfv\'en wave timescale (special case of $\tauW$)\\
        $ \tau_3 (\vk)$ & triple correlation timescale\\
        $ \tauS (\vk) $ & spectral transfer (cascade) timescale:
                          $ \displaystyle \epsilon \approx
                                \frac{ \delta v^2(k)} {\tauS(k)} $ \\
%        $ \vdots $ & \\
      \hline
    \end{tabular}
    \end{center}
  \end{table}
%%              left,center,right.  page 64 of LaTex book;
%       \multicolumn {# cols to cover} { l, c, or r} {text-string}
%       \cline {start_col-end_col}
%       \rule [-1.7ex]{0pt}{5.0ex}
%      \begin {tabular} { || p{5em} | p{8em} ||  }
%------------------------------------------------------------------------------%

% Moreover, GS95 restricted attention to Alfv\'enic fluctuations,
% meaning those polarized like shear Alfv\'en waves,
% namely in the toroidal direction
%         $ \vk \times (\vk \times \vB_0) $.
% The other class of incompressible fluctuations,
% called poloidal or pseudo-Alfv\'en waves,
% were discarded during the development of the CB phenomenology.

%===============================================================

        \section{Antecedents of Critical Balance}
                \label{sec:evolution}

The formulation of critical balance
brought together a number of threads of discussion
of MHD turbulence when a mean magnetic field $\vB_0$ is present.
Important ideas regarding anisotropy in a magnetized plasma,
including the critical balance scenario,
have been developed in part
to understand observations in various systems, including
laboratory confinement devices
        \citep{RobinsonRusbridge71,ZwebenEA79},
the solar wind
        \citep{Coleman68,BelcherDavis71},
and the interstellar medium
        \citep{Higdon84}.
These underlying ideas
included the relative importance of aspects such as
nonlinear activity vs.\ linear wave activity,
incompressible vs.\ compressible fluctuations,
perpendicular vs.\ parallel cascades,
and several kinds of anisotropy with respect to $\vB_0$:
perpendicular and parallel lengthscales,
and
transverse and parallel components of the fluctuations.
Particularly important in the CB context are the
        \emph{timescales}
of the associated processes.

Many of these ideas were well-known.
For example,
there is a long history of
the notion of
  ``Alfv\'enic turbulence''
in the literature,
often spanning somewhat
different meanings.
See
        Appendix~\ref{app:Alfvenic}
for a brief listing of some of these.
As intended here, the idea of Alfv\'enic turbulence
probably originated with the
        Iroshnikov--Kraichnan (IK) phenomenologies
of MHD turbulence
        \citep{Iroshnikov64,Kraichnan65}.
Employing Elsasser variables,
        $\vz^\pm = \vv \pm \vb $,
the MHD equations are
        $ \partial_t \vz^+ \sim - \vz^- \cdot \nabla \vz^+ $,
where only the nonlinear term is indicated, and a symmetric equation
for $\vz^-$ holds.  This form makes it clear that
nonlinear effects require the interaction of fluctuations with
opposite signs of cross helicity,
        i.e., $\vz^+$ and $\vz^-$ both nonzero.\footnote{The
   `opposite signs of cross helicity' requirement for
   nonlinear effects in incompressible MHD is often
   stated in the weaker form `counter-propagating modes are needed.'
   The latter is sufficient, but it is not actually necessary.
   In particular,
        $\vz^\pm(\vx,t) \ne \vct{0} $
   does \emph{not} imply that propagating modes are present, or indeed
   the presence of waves of any kind.}
The IK phenomenologies are based on the
        (weak)
interaction of
counter-propagating Alfv\'en wave packets
in a strong large-scale magnetic field:
        $ \delta v, \delta b \ll B_0 $.
Consequently, the timescale associated with Alfv\'en waves,
        $ \tauA( \vk) = 1 / | \vk \cdot \vB_0| $,
plays a crucial role in these theories---although, tellingly,
its inherently anisotropic nature is neglected with
        $ \vk \cdot \vB_0 \to k B_0 $.
This leads to the well-known IK form for the
omnidirectional energy spectrum,
        $ \Eomni( k) \approx \sqrt{\epsilon B_0} k^{-3/2} $.

Some support for this picture---albeit
        with $\delta b / B_0 \approx 1 $
        rather than small---was
provided by early observations of solar wind fluctuations
        \citep{Coleman66, Coleman67,BelcherDavis71}.
These indicated that,
even though energy spectra typically
had power-law inertial ranges (suggestive of turbulence), the
fluctuations nevertheless had several properties consistent with
large-amplitude  Alfv\'en waves.
For example, low levels of density fluctuations
        (i.e., near incompressibility),
a dominance of polarizations in the plane perpendicular
to the average magnetic field (in a ``5:4:1'' ratio),
and strong correlation of velocity and magnetic fluctuations
(high cross helicity).
Thus the premise that (incompressible) MHD turbulence had wave-like
features seemed reasonable.

The view that low-frequency turbulence\footnote
        {This stands in contrast to \emph{high}-frequency
           ``Alfv\'enic turbulence''
         in the form of so-called
                \emph{weak turbulence}.
         In this the modes act as waves in leading order,
         with weak nonlinear effects accumulating over timescales much
         longer than the wave periods
                \citep[SG94, GS97,][]{GaltierEA00}.
%%                \citep{SridharGoldreich94,GoldreichSridhar97,GaltierEA00}.
        See also \S\ref{sec:tauNL}.}
was the most important kind emerged a little later,
from the experimental context of
disruptions in tokamaks
        \citep[e.g.,][]{Kadomtsev92},
relaxation in RFPs
        \citep{Taylor74},
and
lengthscale anisotropy of the form
        $ \ell_\parallel  \gg  \ell_\perp$,
where these are the correlation
lengths along and across the mean magnetic field
        \citep{RobinsonRusbridge71,ZwebenEA79}.
This prompted development of RMHD
   (aka the Strauss equations),
for which the aim was to obtain simplified equations that retained
nonlinear effects at leading-order,
despite the fluctuations being of small amplitude
relative to the strong mean magnetic field,
  e.g.,
        $\delta b \ll B_0 $
        \citep{KadomtsevPogutse74,RosenbluthEA76,
                Strauss76}.
In deriving RMHD
one eliminates high-frequency wave activity,
so that all
remaining motions are on the ``slow'' advective (or turbulence)
timescale.
Hence any waves present must have timescales no faster than the
nonlinear one:
     i.e., $ \tauW \gtrsim \tauNL $,
a relation that is
an obvious relative of the CB condition,
        Eq.~(\ref{eq:CB-condit}).
Enforcing this timescale inequality necessitates that the fluctuations
have spectral anisotropy of the form
        $ k_z  \ll  k_\perp $.
 Montgomery's (1982)
        \nocite{Mont82-strauss}
derivation of RMHD emphasized these aspects of the physics,
thereby raising awareness regarding two important points:
\begin{enumerate}
  \item[(i)]
        That the physics is different for fluctuations with
             $ \tauNL(k) \lesssim \tauA(\vk) $
         (nonlinear effects crucial)
        versus those for which
             $ \tauNL \gg \tauA $
        (wave dynamics influential);
  \item[(ii)]
      That the anisotropic nature of the Alfv\'en wave timescale,
             $ \tauA( \vk) = 1 / |k_z B_0 | $,
      must be considered.
\end{enumerate}

Spectral anisotropy has entered naturally in the above discussion
but is actually a separate issue,
not originally covered in the Alfv\'enic turbulence categories.
It is now understood however,
that when $\vB_0$ is at least moderately strong, MHD
turbulence evolves towards states with this kind of spectral
anisotropy
        \citep{MontTurner81,ShebalinEA83,Bondeson85,Grappin86,
                CarboneVeltri90,OughtonEA94,
                ChoVishniac00-aniso,MaronGoldreich01,
                BigotEA08-aniso}.
In other words,
perpendicular spectral transfer is strong in these circumstances,
and MHD turbulence consequently lends itself to a low-frequency
description.
This strong perpendicular transfer is an important ingredient in CB
phenomenologies and we discuss it in more detail later.

Building on earlier descriptions of
incompressible MHD turbulence with a strong $\vB_0$
        \citep{MontTurner81,Mont82-strauss},
   \cite{Higdon84}
presented a spectral model for RMHD.
He showed that the RMHD scalings\footnote{Meaning
    those applicable in the classic low plasma beta case.
    Specifically, for a small parameter $\Higdelta $ one has
        $ \delta v_\perp, \delta b_\perp = O(\Higdelta) $
    relative to $B_0 $,
    and
        $ \delta v_\parallel, \delta b_\parallel = O(\Higdelta^2) $
    with
        $ k_z/k_\perp = O(\Higdelta)$.
    However, Higdon goes further and assumes that these scalings apply
    to the IR Fourier amplitudes of
        $ \vv( \vk) $  and  $ \vb( \vk) $,
    with $\Higdelta$  becoming wavenumber dependent:
        $ \Higdelta(k_\perp) = b_\perp(k_\perp) / B_0$.
    Note that Higdon uses $\delta$ in place of $\Higdelta$.}
together with an assumed Kolmogorov IR for the
perpendicular fluctuations
        (e.g.,
                $ E^b_\perp \propto \epsilon^{2/3}k_\perp^{-5/3} $)
imply several things.
These include
a form for the spectra of the parallel components of the fluctuations
        [see his Eq.~(5)],
and, more importantly in the CB context,
the wavenumber relation
 %%     (in our notation)
%-------------------------------------------------------------------------%
 \begin{eqnarray}
       k_z B_0 = A \epsilon^{1/3} k_\perp^{2/3} ,
   \label{eq:Higdon-k}
 \end{eqnarray}
%-------------------------------------------------------------------------%
where $A$ is an order unity
constant.\footnote{This equation
        appears in-line in \cite{Higdon84},
        immediately below his Eq.~(5)
        with the typo $A t^{1/2} $ instead of $\sqrt{A_t}$.}
We refer to this relation as the
(infinite Reynolds number)
        \emph{Higdon curve}.
This appears to be the first time the
        $ k_z  \sim  k_\perp^{2/3} $
scaling was recognised, and it is in fact equivalent to the CB
condition,
        Eq.~(\ref{eq:CB-condit}),
for RMHD:
the LHS is the reciprocal Alfv\'en timescale,
        $ 1 / \tauA( \vk) $,
and the RHS is readily shown to be the (reciprocal) nonlinear
time associated with the assumed perpendicular Kolmogorov
spectrum.
In obtaining
        Eq.~(\ref{eq:Higdon-k}),
it was essential that
(i) the RMHD small parameter  $ \Higdelta $
was extended to be  $ k_\perp $-dependent:
        $ \Higdelta (k_\perp) = b_\perp(k_\perp) / B_0 $,
and
(ii) that
        $ k_z / k_\perp$
was set {equal}
to        $ \Higdelta (k_\perp) $,
not just of order    $ \Higdelta$.
Here,
        $ b_\perp( k_\perp) = \sqrt{ k_\perp E^b_\perp(k_\perp)} $
is the rms strength of the perpendicular magnetic field fluctuations
at perpendicular scales
        $ \sim 1/k_\perp$.

It is also noteworthy that Higdon
        (1984, \S III)
%%        \cite[\S III, p.~112]{Higdon84}
was emphatic that his proposed model of turbulence could
        \emph{not}
  \emph{``be interpreted in the context of nonlinear analogs of the
          linear characteristic mode of MHD: \emph{propagating}
          Alfv\'en waves.''}
Indeed, he also remarked
%%        (in the same paragraph)
that although turbulent fluctuations may sometimes be identified with
strongly interacting nonlinear
        \emph{analogs}
of disturbances satisfying the linearized equations,
the
   \emph{``nonlinear variations possess fundamental properties not
           found in linear modes. Their non-modal nature is essential
           to the existence of turbulent cascades.''}
Evidently
he was not viewing (R)MHD turbulence as the
interaction of wave-like fluctuations,
whereas that is the perspective
in the IK phenomenologies.

Around the same time as
        Higdon's
        (\citeyear{Higdon84})
work appeared,
        \cite{ShebalinEA83}
proposed a weak turbulence explanation for the dominance of
perpendicular spectral transfer.
This treats the leading-order fluctuations as linear Alfv\'en waves and
uses perturbation theory to calculate nonlinear corrections.
The most important corrections are due to so-called
        ``3-wave resonant interactions.''
However, as is now well-known, in MHD one of these modes is not
actually a wave at all,
but rather a non-propagating 2D      ($k_z = 0 $) mode
whose role is to mediate the transfer of energy between two waves
that are propagating in the same direction and have the same $k_z$
        \citep{ShebalinEA83,Bondeson85,Grappin86}.
Naturally, the cross helicity of the 2D mode must be opposite in sign
to that of the two wave modes.
The crucial point is that the first nonlinear correction
involves transfer of energy at fixed $k_z$---i.e., a strictly
perpendicular cascade.
See
        Appendix~\ref{app:Shebalin}
and the original papers for details.\footnote{A complete \emph{strong}
        turbulence explanation for strong perpendicular spectral
        transfer is still being sought, although the pathway by which
        it occurs has been identified in the context of the von
        K\'arm\'an--Howarth correlation equation hierarchy
                \citep{WanEA12-vKH,OughtonEA13}.
        See also \S\ref{sec:strong-frustrated-iso}.}

Summarizing,
the theories and models of MHD turbulence discussed in this
section---IK,
  Shebalin et al.'s  perpendicular transfer,
  and RMHD (including Higdon's model)---all assume
small amplitude fluctuations relative to a strong
large-scale magnetic field.
However, they are not all strong turbulence models.
The IK phenomenologies are weak turbulence ones since
they assume that the nonlinear timescale is long compared to the wave
timescale and hence that the spectral transfer time, $\tauS$, is also long.
The
        \cite{ShebalinEA83}
explanation for strong perpendicular transfer is based on weak
turbulence features.
  Higdon's (\citeyear{Higdon84})
spectral model for RMHD is a strong turbulence approach,
in the sense that there are no timescales faster than the nonlinear one,
        $\tauNL$.
As noted in GS95, with the benefit of hindsight it is clear that CB
ideas are incipient in Higdon's approach,
and in the RMHD derivation presented by
        \cite{Mont82-strauss}.

This completes our discussion of the antecedents of CB.  Before
proceeding to an examination of its original presentation
%% and subsequent development,
we first
discuss how the nonlinear timescale impacts the weak or strong nature
of turbulence.

%===============================================================

%\clearpage
%%        \section{Timescales and Weak vs.\ Strong turbulence}
        \section{$\tauNL$ and Weak vs.\ Strong Turbulence}
                \label{sec:tauNL}

In the previous section we have seen that it is important to
distinguish between weak turbulence and strong turbulence, and that
this can be accomplished via comparison of the scale-dependent
nonlinear and wave timescales.
This is also an important issue in CB contexts.
In this section we provide definitions for these terms, employing
incompressible 3D MHD (with a $\vB_0$) as a
        representative example.
The wave timescale is then the Alfv\'en one,
        $ \tauA( k_z) = 1 / | k B_0 \cos \theta | $,
and is clearly anisotropic
        ($ \theta $
         is the angle between $\vk$ and $\vB_0 $).
% We begin by defining weak and strong turbulence since there are
% subtleties associated with definition of $ \tauNL( k) $.

A necessary condition for weak (aka wave) turbulence is that the
fluctuations are of small amplitude, since otherwise nonlinear effects
would be present at leading order.\footnote{
        A very reasonable definition of turbulence might include the
        requirement that nonlinear effects are present at leading-order.
        The term `weak turbulence' would then be inappropriate
        since for it, nonlinear effects are higher order corrections that
        accumulate over long times.  Nonetheless, usage of this
        terminology is well-established.}
However, as the RMHD model reveals, this is not a sufficient
condition.
To develop a more complete definition,
consider a set of fluctuations with wavevectors $\vk$
near some chosen value $ \vk_a $.
When, for all these nearby fluctuations,
          $ \tauNL( k_a)  \gg  \tauA( \vk) $,
they are said to be weakly turbulent,
where the nonlinear timescale is defined using the whole shell
        $ k \approx |\vk_a| $
(see below).
The point being that the wave timescale
is much faster than the nonlinear one.
If instead
          $ \tauNL( k_a)  \lesssim  \tauA( \vk)$
for these fluctuations, they are called strongly turbulent.
This includes situations where the nonlinear time is much shorter
than the wave timescale.
The idea is that any wave effects are likely to operate too slowly to
%%    overcome or
   substantially disrupt
the nonlinear processes.

If one of these timescale inequalities holds for the whole system, one
speaks of turbulence that is
        \emph{globally}
strong or weak.\footnote{In both cases, one needs a substantial range
        of scales where dissipation effects are negligible.  This is
        tantamount to having large Reynolds numbers.}
GS95 have shown that
an initial state of globally weak turbulence is often unstable,
with strongly turbulent fluctuations developing at small perpendicular
scales.
This is discussed further in    \S\ref{sec:weak-to-strong},
along with subcategories of weak turbulence that depend on whether 2D
($k_z = 0$) modes are excited or not
        (GS97).

How is the nonlinear timescale defined?
Recall that $\tauNL$
arises from the bracketed part of the
        $ (\vv \cdot \nabla) \vv $
term in the momentum equation
        \citep[e.g.,][]{Frisch}.
Thus in a global sense one has
        $ \tauNL \approx   L / \delta v $,
where   $ L $
is a characteristic (`energy-containing') scale for the rms velocity
fluctuation $\delta v$.
This estimate is appropriate for isotropic turbulence, but will
usually require refinement for anisotropic systems,
including MHD with a mean field.
For example, when the turbulence is anisotropic at the
energy-containing scales, a single
        $ L $
is insufficient to characterize this range.
Knowledge of dynamical tendencies---such as
spectral transfer that is predominantly perpendicular---can be used
to provide better estimates for the nonlinear time,
        e.g., $ \tauNL = L_\perp / \delta v $.

A definition for the nonlinear timescale associated with
IR scales is also needed. In this regard a standard viewpoint
is that nonlinear couplings are predominately
local in scale. This is central in the
        \cite{Kol41a}
theory and is equivalent to the statement that the nonlinear
timescale at wavenumber $k$ depends
only on $k$ and the turbulence amplitude at $k$.
For isotropic turbulence,
one simply generalizes the global form to
        $\tauNL( k) = 1 / [k \, \delta v(k)] $,
where
        $ \delta v(k) $
is an estimate for the mean speed at scales
        $ \ell  \approx   1 / |\vk| $.
This can be calculated in several ways.
The omni-directional
energy spectrum,\footnote{Recall that for a Navier--Stokes fluid,
                Kolmogorov phenomenology yields a
                powerlaw IR spectrum,
                   $ \Eomni( k) \approx  \epsilon^{2/3} k^{-5/3} $.}
  $ \Eomni( k) $,
can be employed
with
        $ \delta v(k)^2  \approx  k \Eomni( k) $,
yielding
\begin{equation}
         \tauNL (k) = \frac{1} { k \sqrt{ k \Eomni( k)} }
  .
   \label{eq:tauNL}
\end{equation}
Alternatively,
an estimate can be constructed in coordinate space, using the
mean-square velocity difference across pairs of points with relative
separation    $ \ell $:
        $ \delta v_\ell^2
            =
        \avg{ |\vv( \vx) - \vv( \vx + \ehat \ell)|^2 } /2 $,
where
        $ \ehat$
is a unit vector and the angle brackets denote averaging over
        $ \vx $.
Often, averaging over directions of $\ehat$ is also
included.\footnote{The
        normalization factor of a half ensures that when
          $ \ell  >  L_\text{cor} $,
          $ \delta v_\ell \to \delta v $,
        the global rms fluctuation strength.}

Note that these definitions are based on the rms speed for the
        \emph{entire shell}
of wavevectors with magnitudes close to $ | \vk | $,
and not on a speed associated with wavevectors close in solid angle
(and magnitude) to $ \vk $.
In other words, the rms speed is local in $k$-magnitude,
i.e., local in scale,
but includes contributions that are nonlocal with respect to the
direction of $\vk$.
This would certainly be an appropriate choice for isotropic
turbulence, but the issue become more subtle for
anisotropic MHD
%% Further discussion is available in the literature
        \citep[see, e.g.,][]{ZhouEA04,MattEA09-kspdiff}.

\begin{figure} [tbp]
 \begin{center}
    \includegraphics[width=\columnwidth]{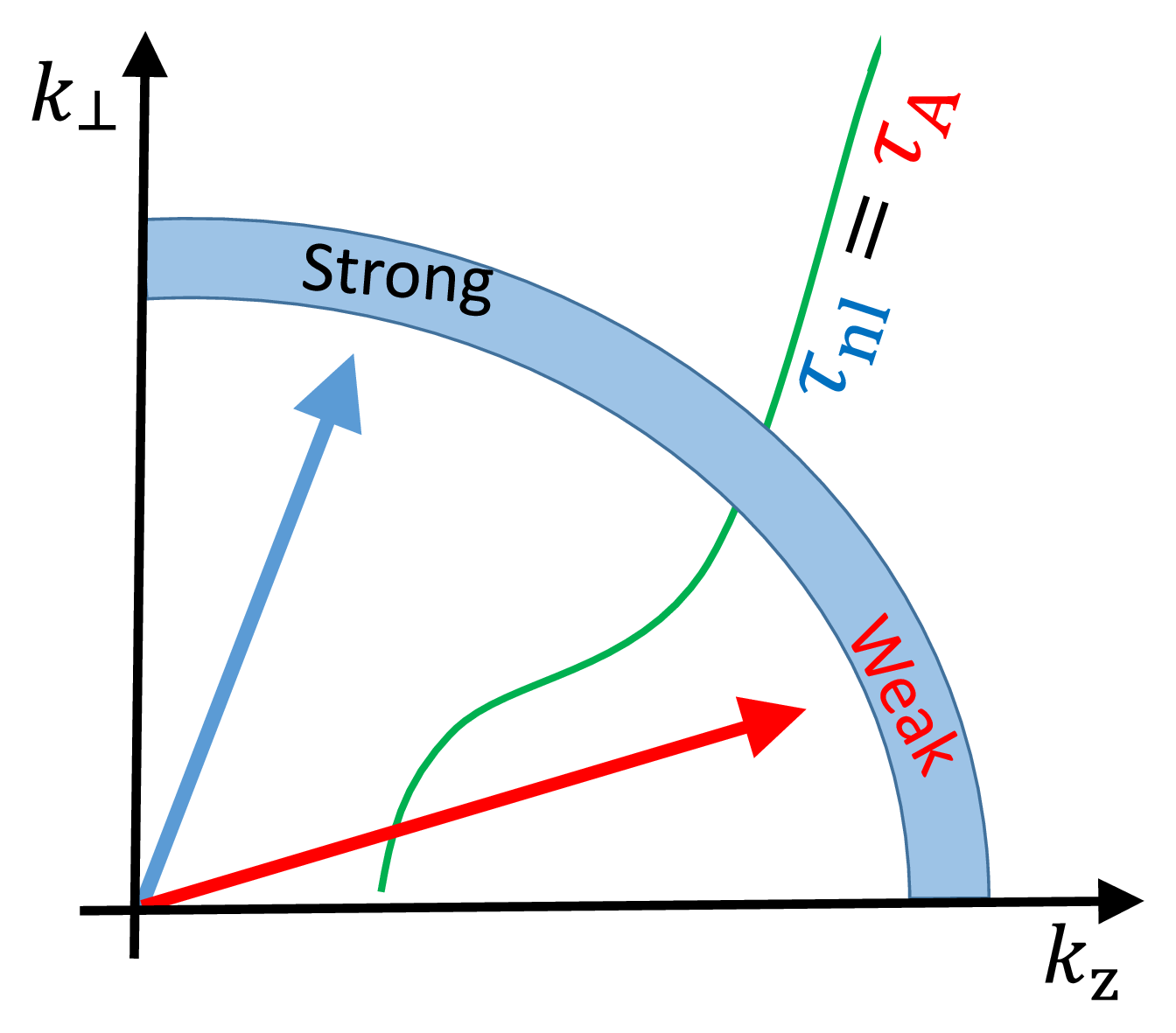}
 \end{center}
  \caption{Indication of how excited modes on a spherical shell in
        $\vk$-space
    can be either weakly turbulent or strongly turbulent depending
    upon the orientation of their wavevector
        (relative to the equal timescale curve and  $B_0 \zhat$).
          }
  \label{fig-weakVstrong}
\end{figure}

The anisotropy of $\tauA = 1 / | \vk \cdot \vB_0| $
means that
on a spherical shell of radius $ |\vk | $,
this wave timescale formally varies between
     $ 1/|k B_0| $
and infinity.
It follows that knowing the scale of the fluctuation,
         $ \ell  \approx  1 / |\vk| $,
is insufficient to determine whether the fluctuations near $\vk$ are
weak or strong.
For example, on the same spherical shell in $\vk$-space, it is quite
possible to have regions where the turbulence is strong, and others
where it is weak, as indicated in
        Figure~\ref{fig-weakVstrong}.
This depicts a sample energy spectrum for which the excitation
outside    (i.e., at larger $k_z$)
the equal timescale curve is very low.
Clearly, the fluctuations will be
weakly turbulent near the region labelled `Weak'
and strongly turbulent over most of the rest of the shell.
%-----------------------------------------------------------------------------%
%-----------------------------------------------------------------------------%

There is one further point to make regarding $\tauNL( k)$ for
incompressible MHD.
When all the $\vv(\vk)$ and $\vb(\vk)$ fluctuations are toroidal,
  i.e., polarized parallel to
           $ \vk \times \vB_0 $,
one sees that
        $ \vv \cdot \nabla   \to  \vv \cdot \vk \approx k_\perp v $.
This is the situation
whenever        `Alfv\'en mode turbulence'
is considered, and in particular,
this is the case treated by GS95.
An appropriate definition for the nonlinear time associated with
toroidal (Alfv\'enic) fluctuations
is then
\begin{equation}
         \tauNL^\perp( k) =  \frac{1} { k_\perp \sqrt{ k \Eomni( k)} }.
  \label{eq:tauNL-tor}
\end{equation}
As this is a function of $k_\perp$ and $k$,
it would be correct to write
        $ \tauNL^\perp $
with a vector $\vk$ argument, rather than the scalar $k$
we have been using.
We elect not to do so herein, in order to emphasize the
   `shell based'
nature of the estimate for      $\delta v (k) $
employed in     $ \tauNL^\perp $.

We note that much of the
above discussion is readily extended to the case of
nonzero cross helicity
        \citep{GrappinEA82,PouquetEA86,HossainEA95,Boldyrev06,WanEA12-vKH}.
However, this will not be a central theme in this review.

%===============================================================

        \section{Critical balance: GS95 Derivation}
                \label{sec:emergence}

Having reviewed the relevant work occurring prior to the emergence of
CB, we are now ready to consider the original (GS95) derivation
of critical balance theory.
The key assumptions made in the GS95
description of MHD turbulence are:
\begin{itemize}
  \item that the system of interest is
  incompressible 3D MHD with a uniform mean magnetic field
        $ \vB_0 = B_0 \zhat$;

  \item that the dynamics of interest consists of nonlinear
    interaction of waves,\footnote{For the strong turbulence
      case, GS95 note (their footnote 2) that:
     \emph{``interactions are so strong that a  `wave packet'
         lasts for at most a  few wave periods,''}
        so that the integrity of the wave properties is unclear or
        perhaps even strongly lacking.
        }
    which must be Alfv\'en waves
    (to satisfy incompressibility);
  and

  \item that the appropriate basis is that of linear eigenmodes,
    so that only toroidally polarized fluctuations are considered
    (poloidal ones are discarded).
\end{itemize}

\begin{figure} [tbp]
 \begin{center}
   \includegraphics[width=\columnwidth]{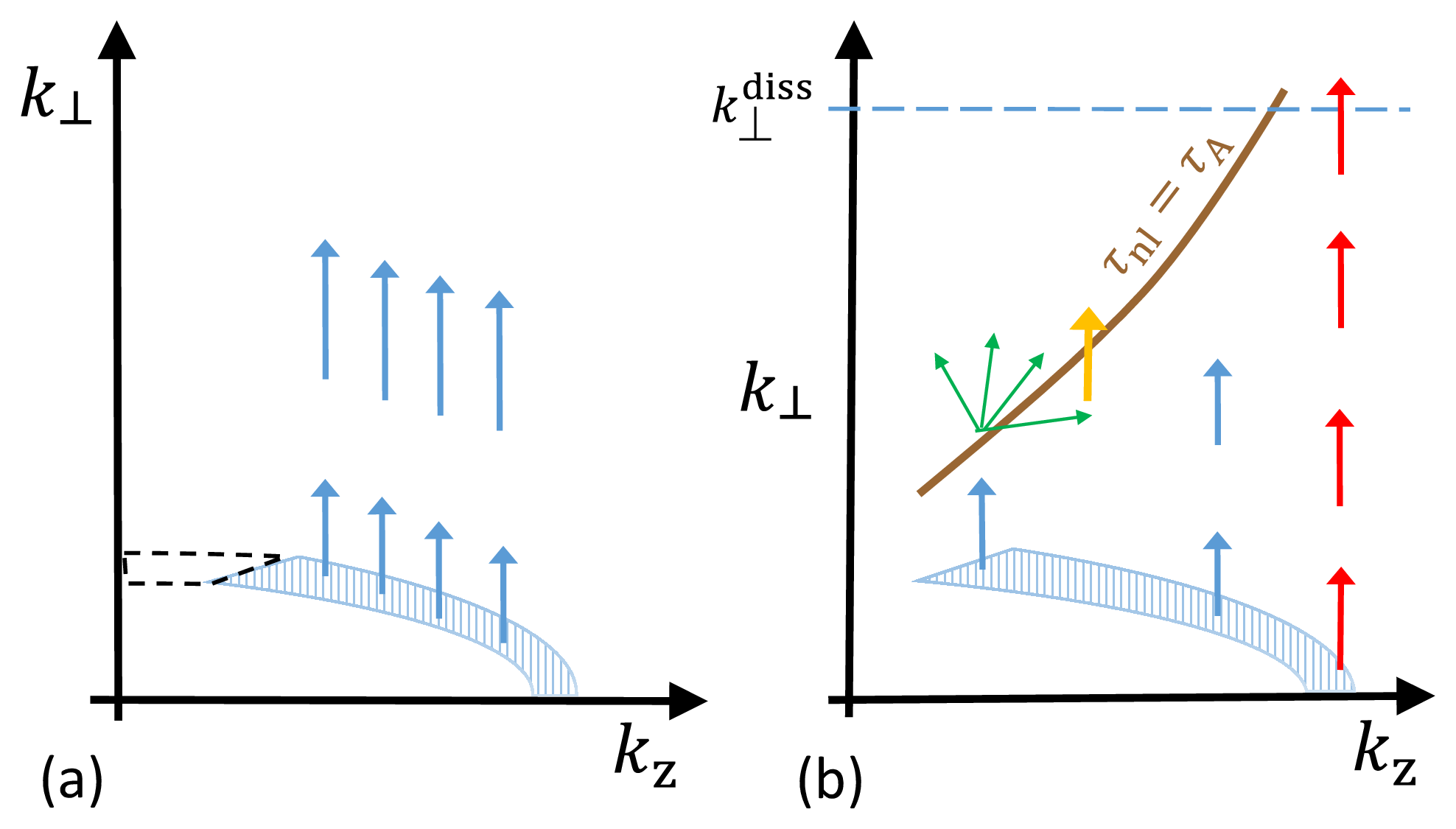}
%%    \includegraphics[angle=90,width=\textwidth]{XXX}
%--------------------------------------------------------------------------------
 \end{center}
  \caption{(a) Schematic spectrum for sample initial weak turbulence
              state that is roughly isotropic  (shaded area),
              except for exclusion of 2D modes (empty dashed zone).
              Blue arrows show dominant direction of spectral transfer
              for weak turbulence.
           (b) Spectrum at a later time showing how initial
              perpendicular transfer leads to establishment of an
              equal timescale curve (brown line) and CB zone around it.
              Energy from the weak turbulence fluctuations continues
              transferring to larger $k_\perp$   (blue arrows)
              until the associated      $\vk$  has a
                 $\tauNL \approx \tauA $.
              At such positions the energy (now green) transfers
              approximately isotropically.  Energy that moves outside
              the equal timescale curve is back  in a weak turbulence
              region and again subject to strong perpendicular
              transfer (gold arrow).
              Note that energy that starts at large enough $k_z$
                (red arrows),
              arrives at the dissipation scale, $k_\perp^\text{diss}$,
              without encountering the equal timescale zone.
          }
  \label{fig-CBweak}
\end{figure}

GS95 lay out two scenarios in which CB can be relevant (their \S2),
which are discussed below in greater detail.
In the first of these,
   $B_0$ is strong
and there is a weak turbulence state in which CB does not hold initially,
but rather develops at some small scales.
For this part of the spectrum,
GS95 envisions that four-wave weak turbulence couplings
lead to transfer to higher $k_\perp$
without increasing $k_\parallel$.
Thus a highly anisotropic state emerges
with
        $ k_\perp \gg k_\parallel $.
In the second scenario,
the energy-containing scales are assumed to be
critically balanced from the outset,
with subsequent dynamical population of IR scales also occurring in
accord with CB.
In both cases the energy-containing (large) scales are assumed
to be more or less
isotropic.\footnote{In this context the GS95 definition of
        isotropic is that the characteristic parallel and
        perpendicular scales are comparable.
        However, it apparently does not imply a uniform
        distribution of power over all wavevector directions
        since low-frequency (small $k_z$) fluctuations seem
        to be excluded.
        This definition is a non-standard one compared to the usual
        concept of `independent of angles' used in turbulence work.
        }
The former case applies for strong mean magnetic field,
        $ \delta B/B_0 \ll 1 $,
and the timescales, evaluated at the
outer scale, ordered so that
        $ \tauA < \tauNL $
   (cf.\ Fig.~\ref{fig-CBweak}).
The latter case for isotropic outer scale fluctuations,
is only feasible
if      $ \delta b  \sim B_0$
so that the wave and
nonlinear timescales might be equal.

%% As we have stated,
A key step in GS95 is to compare the
nonlinear and wave timescales.
The GS95 estimates of these
are
        $ \tauNL^\perp( k) = 1 / [ k_\perp \delta v(k) ] $
and
        $ \tauA (\vk) = 1 / | k_z B_0 | $;
see \S\ref{sec:tauNL}.
Based on these assumptions and approaches, the
steady turbulence phenomenology known as
critical balance emerges,
as we now explore.

%-----------------------------------------------------------------------------%
%-----------------------------------------------------------------------------%

    \vfill

        \subsection{Scenario 1: Weak to Strong Transition}
                \label{sec:weak-to-strong}

The initial state considered
has a \emph{strong} mean field  $ B_0$
with an
        (almost)
isotropic spectrum of small-amplitude Alfv\'en waves.
The latter are subject to weak nonlinear interactions
%\footnote{The
%                SG94 and GS95
%        definitions of weak turbulence \emph{require} that there is no
%        energy in the
%                $ k_z = 0 $ (aka ``zero frequency'' or 2D)
%        modes, so that the leading-order nonlinear interactions are
%        4-wave ones.
%        A more general weak turbulence case has the 2D modes
%        excited (occasionally called ``intermediate turbulence'')
%        and thus supports 3-mode interactions, and indeed these are
%        the leading-order nonlinear couplings
%            \citep{ShebalinEA83,GoldreichSridhar97,
%                   GaltierEA00,NazarenkoEA01,Galtier}.
%        Moreover, even when 2D modes are initially absent, 4-wave weak
%        turbulence couplings will often immediately generate 2D modes,
%        and these then engage in 3-mode couplings.
%        See later in this section.}
and so satisfy
        $ \tauA(\vk) \ll \tauNL^\perp( k) $
for each of the excited wavevectors $ \vk$.
This is frequently an unstable situation, transitioning
to a state that is strongly turbulent at large $ k_\perp $.
Dynamically,
the important point is that the weak turbulence
cascade is dominantly towards higher perpendicular wavenumbers,
and therefore transfers
energy from fluctuations for
which
        $ \tauA \ll \tauNL^\perp $
to ones for which
        $ \tauA \approx \tauNL^\perp $.
That is, the transfer is towards a CB state.
See
        Figure~\ref{fig-CBweak}.
The GS95 argument is based on a full acceptance of the reasoning
from their earlier paper on weak four-wave turbulence
        (SG94).
%%        \citep{SridharGoldreich94}.
   % This leads to anisotropic spectral transfer towards higher
   %     $ k_\perp $
  % at essentially fixed $ k_z $,

For emphasis,
we summarize the several stages of the GS95 reasoning for scenario 1 as:
%   \item weak fluctuations:
%         $\tauNL(\vk) \gg \tauA(k_z)$; typically the large
%         $k_z$ case.
    \begin{enumerate}
      \item[1)]
     Weak turbulence 4-wave interactions produce spectral transfer
     that is dominantly towards larger $k_\perp$, at essentially fixed
        $ k_z$.

      \item[2)]
     Because the nonlinear time
        $ \tauNL^\perp(k) $     %% = 1 / [k_\perp \delta v(k)] $
     typically decreases with increasing $ k_\perp$,
     this transfer
     means the higher-$k_\perp$ fluctuations have \emph{stronger}
     nonlinear interactions.
%%     causes the initially weak nonlinear interactions to strengthen.

      \item[3)]
     As the transfer continues, the regions of spectral space for
     which
        $\tauNL^\perp(k) \approx \tauA( \vk) $
     become significantly populated;
     that is, those fluctuations are in a state of critical balance.

      \item[4)]
     At still larger $k_\perp$, the spectral transfer occurs in such a
     way as to maintain the CB condition.
%%         $ \tauNL( k) \approx \tauA( \vk) $.
     Because the latter implies a relation between $k_z$ and
     $k_\perp$, some parallel transfer also occurs.
    \end{enumerate}

These four points in effect paraphrase the
three points emphasized by
GS95 when they ``take stock of the arguments'' leading to CB
        (their page 764).
We note that GS95 also
stress that
  ``each of the these statements is based on weak 4-wave couplings.''
In this scenario
excitation becomes concentrated along and `near' the equal timescale
curve, producing a ridge-like IR spectrum (of undetermined width)
for the scales that are critically balanced.
Note that
the $k_z = 0 $, or 2D, modes are not discussed in GS95, but their
status is implied:
specifically, in order that the underlying
4-wave couplings control the spectral  transfer as stated,
the 2D modes
must remain unexcited, or, perhaps,
negligibly excited.

%----------------------------------------------------

        \subsubsection{Commentary.}
                \label{sec:weaktostrong-comm}

There are several points to discuss regarding this case,
wherein CB
develops dynamically, but was not present initially.
These include the absence of 2D modes,
perpendicular spectral transfer,
and the dynamics for large values of  $ k_z $.
These are now considered.

   \paragraph{Absence of 2D modes.}%
%        \label{sec:weak-no2D}
Various issues are associated with the
assumed absence of these fluctuations.
The first thing to note is that if 2D (or $\approx$ 2D) modes are
excluded from the initial state,
the fluctuations are not properly isotropic since the 2D
wavevectors then constitute distinguished directions.
So why would one wish to exclude them?
As discussed in    \S\ref{sec:tauNL},
for a spherical shell in $\vk$-space with roughly isotropic
excitation, there are always some modes that cannot be weakly
turbulent, because of the anisotropic nature of
the Alfv\'en wave timescale
        $ \tauA = 1 / |k B_0 \cos \theta| $.
Thus, if one seeks an initial state in which all
fluctuations are weakly turbulent (not just small amplitude), one must
exclude at least the 2D and quasi-2D modes;
by definition,
these modes
satisfy the strong turbulence criteria of
        $ \tauNL(k) \lesssim \tauA(\vk) $, for which the
  \emph{nonlinearity parameter}\footnote{Denoted as
         $ \zeta_\lambda = \tauA(\vk) / \tauNL^\perp(k) $
    in GS95 and $\chi$ in many subsequent works, including this one.
         $ \chi$
    may also be defined in terms of the timescales at a specified
    spatial lag.}
        $ \chi = \tauA / \tauNL $
can be arbitrarily large.
See
        Figures~\ref{fig-weakVstrong} and \ref{fig-CBweak}.
If they are not excluded, the initial state is likely to involve
RMHD fluctuations
        (discussed in \S\ref{sec:rmhd}).

A follow-up paper
        (GS97)
to GS95 acknowledges that there is also
an  ``intermediate turbulence'' case, in which the 2D modes
        \emph{are}
present in the initial state.
However, by imposing an assumption
of small amplitude,
       $ \delta v, \delta b \ll B_0 $,
it is possible to recover
again a weak turbulence situation
(ignoring the possibility of RMHD fluctuations).
It is argued that all orders contribute equally in the perturbative expansion,
but this has since been shown to be incorrect
        \citep{NazarenkoEA01,LithwickGoldreich03}.
The leading-order weak turbulence situation, with 2D modes excited,
has been considered in detail
        \citep{GaltierEA00,GaltierEA02,Nazarenko}.
Of course, the self-interaction of the 2D modes is not necessarily
weak, and in general could be strongly turbulent.
For example, in the solar corona, some types of random motions of
the photospheric footpoints of the magnetic fieldlines will produce
strongly turbulent 2D modes
     \citep[e.g.,][]{DmitrukGomez97,DmitrukGomez99,
                DmitrukEA01-quebec,DmitrukMatt03,
                RappazzoEA08,RappazzoEA10}.
Such cases are often related to RMHD with its inherent
requirement that 2D and quasi-2D modes are strongly turbulent.
See     \S\ref{sec:rmhd}.

        \paragraph{Perpendicular transfer.}
While the idea of dominant perpendicular spectral transfer is
correct for weak turbulence modes, it is unusual for this to be due to
4-wave interactions.  Rather, the usual mechanism is the 3-mode process
reviewed in
        \S\ref{sec:evolution} and Appendix~\ref{app:Shebalin}.
This shortcoming of GS95 was quickly addressed
      \citep[GS97]{MontMatt95,NgBhattacharjee96}.
In order
for the 4-wave process to be dominant, there
needs to be a low-$k_z$ cutoff in the
   %% ($v$ and $b$)
energy spectra
        (GS97).
In particular, the
        2D  ($ k_z = 0 $),
modes must be unexcited (aka ``empty''), and likewise the quasi-2D
modes.

However, 2D and quasi-2D modes typically will be
present, unless the boundary conditions prohibit them
     \citep{MontMatt95,GoldreichSridhar97,DmitrukEA01-quebec}.
Moreover, even when 2D modes are absent from the initial state,
the interaction of
counter-propagating Alfv\'en waves\footnote{Or,
        more generally, the interaction of fluctuations whose $k_z$
        are equal in magnitude but opposite in sign.}
with the same
        $ | k_z | $
immediately generates
        $ k_z = 0 $
excitation, when the boundary conditions permit this.
This has been discussed in the literature many times, with
analytic, experimental, and simulation support presented
        \citep[e.g.,][]{NgBhattacharjee96,
                        VasquezHollweg04-jgr,VasquezEA04,
                        HowesNielson13,NielsonEA13,DrakeEA13}.
Quasi-2D modes can be produced in similar fashion.

Once generated, 2D modes (and quasi-2D modes)
are immediately available to play their
role in the 3-mode resonant perpendicular transfer process
        \citep{ShebalinEA83,Grappin86,OughtonEA94}.
In this situation, the perpendicular transfer occurs as a consequence
of successive 3-mode couplings that are of distinct types.
The first involves
parallel spectral transfer and excitation of
        $ k_z = 0 $
fluctuations.
Whereas, in the second type,
the 2D modes and the propagating
modes interact to produce
perpendicular transfer, but no parallel transfer
      \citep{VasquezHollweg04-jgr,VasquezEA04,HowesNielson13}.
Since
        $|\omega| = | k_z V_A | $
for Alfv\'en waves,
this latter class of coupling is described as occurring at
constant frequency.

Such generation of 2D modes\footnote{Called
        ``replenishment'' in the two-component model of
        \cite{OughtonEA06-2cpt}.}
is also
relevant to strong turbulence cases that
initially lack excitation of $k_z = 0 $ modes
        (see discussion in next section).
The coupling that produces this transfer
is \emph{non-resonant}, as is any incompressible transfer that
involves an $O(1)$
change in wave frequency.
We may allow of course for
resonance broadening and quasi-2D couplings
that result in small
frequency changes on the order of $1/\tauNL$.

Although the wave-wave interaction that generates 2D modes is a
non-resonant
process,\footnote{Meaning that although
        the two driving waves are
        solutions of the linearized equations,
        the mode that is driven is not: it is a nonlinear mode
          \citep[e.g.,][]{HowesNielson13}.}
and can be rather weak when $k_z$ is large,
this has little qualitative impact on the occurrence of perpendicular
transfer.
It is clear from the above discussion
that in weak turbulence
the major role of a 2D fluctuation is to couple with a propagating
Alfv\'en mode (with some $k_z$)
to drive another propagating mode, with that same $k_z$
        \citep{ShebalinEA83,Bondeson85,Grappin86}.
In that 3-mode process, the energy transfer is between the two
propagating modes and occurs at fixed $k_z$.  The energy of the 2D
mode is unchanged and it can be considered a mediator or catalyst mode.
The amplitude of the 2D mode affects the
        \emph{rate}
of perpendicular spectral transfer, but not the amount of energy
available for such transfer.

        \paragraph{Behavior at large   $ k_z $.}
As indicated via the red arrows in
        Figure~\ref{fig-CBweak}(b),
this can involve perpendicular transfer that encounters the
dissipation scale without meeting the CB curve, so that what we have
labelled Stage~3
in      \S\ref{sec:weak-to-strong}
does not eventuate.  For these fluctuations,
perpendicular transfer acts to
        move
the energy to the perpendicular dissipation scales,
while the nonlinearity parameter
        $ \chi = \tauA / \tauNL^\perp $
remains less than unity throughout.
Thus these modes act as
weak turbulence, whether the transfer is of the 3-mode type
(2D modes present) or 4-wave type (with 2D modes absent).
This large $k_z$ situation is the one considered by
        \cite{GaltierEA00}.
There is no discussion of CB therein, because it
does not occur in this spectral range.
When the perpendicular dissipation wavenumber,
        $ k_\perp^\text{diss} $,
can be estimated,
the CB wavenumber relation for incompressible MHD,
        $ k_z \sim k_\perp^{2/3} $
    [see    Eq.~(\ref{eq:kz-kperp})],
can be employed to determine a minimum $k_z$ above which this
        `always weak'
cascade occurs
  (see later sections).
The effect of this dissipation range cutoff
on observable 1D reduced solar wind spectra
is discussed in some detail by
	\cite{TesseinEA09}.
%.. See their eqn 6 and after

        \paragraph{Why does the cascade preserve CB?}
A final discussion point relates to
what we have labelled Stage 4 in the above list for the stages of
scenario 1:
why, once CB is established over some scale range, does the
        $ \vk $-space
dynamics maintain this property as it moves excitation to
smaller scales?
This interesting piece of physics is taken up (late) in the next section.

%-----------------------------------------------

        \subsection{Scenario 2: Strong Turbulence}
                \label{sec:CB-strong}

The second CB scenario GS95 consider---and their primary focus---is
that of steady-state \emph{strong Alfv\'enic turbulence},
which they define as follows.
Fluctuations at the
global scales, $ \sim L $,
are assumed to
(i) be roughly isotropic,
and
(ii) have amplitudes comparable to that of the mean field:
        $ \delta v_L, \delta b_L  \sim  B_0$.
Note that
in contrast to the weak turbulence and RMHD situations,
        $B_0$
        is \emph{not} large.
Again, all fluctuations are assumed to be toroidally polarized, i.e.,
in the same sense as linear Alfv\'en modes.
These requirements imply
        $ \tauA  \approx  \tauNL^\perp $
at the global scales, or equivalently
that CB holds for the energy-containing scales.
% Thus, as a consequence of the GS95 definition of strong turbulence,
% CB is an inherent feature of the isotropic energy-containing scales.

To develop the strong turbulence phenomenology
GS95 present two approaches.
First a heuristic discussion is given,
based mainly on weak turbulence reasoning.
The transverse (solenoidal)
property of the linear Alfv\'en mode is crucial in this
discussion.
In a second approach,
GS95 carry out an EDQNMA closure calculation \citep{Orszag70b}
employing toroidal fluctuations, that is,
a representation in which there are
no fluctuation variances in the direction of the global mean
magnetic field.
GS95 state that this restriction is
largely a ``guess'', justified in
part based on the
theory of compressive wave damping
in linear Vlasov theory.

Since this is strong turbulence, the (cascade) dynamics
moves energy from the energy-containing scales to smaller (IR) scales.
As for the weak-to-strong scenario,
this is argued to occur in such a way that CB also ensues at the
smaller scales.  GS95 view that the wave
properties of the fluctuations are important in this cascade
but the waves are not long-lived, stating that
the
     \emph{``interactions are so strong that a  `wave packet'
         lasts for at most a  few wave periods''}
(their footnote 2).

Given this landscape, GS95 develop a functional form for the IR
spectrum of critical balance strong turbulence.
This requires further assumptions.
The first is the
familiar
approximation of
steady-state local-in-scale transfer
in which
the rate of energy injection at the energy-containing scales,
        $ \epsilon_\text{inj}  \approx  \delta v_L^3 / L $,
is equal to the energy cascade rate for inertial range scales,
        $ \epsilon  =  \delta v( k)^2 / \tauS(k) $,
 where
        $\tauS$
 is the spectral transfer (aka cascade) timescale.
As the fluctuations are assumed to be Alfv\'enic the nonlinear
timescale is
         $ \tauNL  \approx   1 / [k_\perp \delta v(\vk)]
                = \tauNL^\perp $.
At this point CB is invoked and
        $ \tauS(k) $
is replaced\footnote{More accurately,
        $ \tauS $
   should be calculated from the relationship
       $ \tauS \tau_3 = \tauNL^2 $
  with $ 1/\tau_3 = 1/\tauNL + 1/\tauA $
       \citep{Kraichnan65,PouquetEA76,MattZhou89-IR,Frisch,ZhouEA04}.
   However, because CB is assumed to hold this only introduces
   a factor of two.}
by
        $ \tauNL^\perp(k) \equiv \tauA(k_z) $.
Using this equivalence in
        $ \epsilon_\text{inj} = \epsilon $
yields
\begin{equation}
      \frac {\delta v(k) } {\delta v_L }
          =
      ( k_\perp L )^{-1/3}
      =
\frac {\delta v(k) } {B_0}
 .
  \label{delta-v}
\end{equation}
The left-hand equality is more general
while the right-hand one gives the
relation in the form written in GS95
and is specific to
their strong turbulence requirement that
        $ \delta v_L \sim B_0 $
 [see their Eq.~(5)].
Although formally correct for the conditions assumed,
this is easy to misinterpret since it suggests, wrongly, that
        $ \delta v(k) $
scales with $ B_0 $---independently of the global
turbulence amplitude
        $ \delta v_L $.
We view
the first equality in  Eq.~(\ref{delta-v})
as a more physically consistent way to express
the scaling.\footnote{See
                \cite{ChoVishniac00-aniso}
                for the equivalent relations that follow from using
                $ \epsilon_\text{inj}  \approx  \delta v_L^2 / (L/B_0) $.}
Alternatively, one can avoid reference to either
        $ \delta v_L $  or  $ B_0 $
by using
        $ \delta v(k)  =  ( \epsilon_\text{inj} / k_\perp )^{1/3} $.

Substituting
        Eq.~(\ref{delta-v})
into the Alfv\'enic IR CB condition,
        $ \tauA(\vk) = \tauNL^\perp(k) $,
leads to the wavenumber relation
\begin{equation}
      k_z  =  \frac{\delta v_L} {B_0} \, k_\perp^{2/3} L^{-1/3} .
  \label{eq:kz-kperp}
\end{equation}
%% which indicates that $k_\perp $ and $k_z$ are correlated.
This is Eq.~(4) in GS95, although they omit the factor
        $ \delta v_L / B_0 $
because it is $ O(1) $
for their definition of strong turbulence.
GS95 interpret the above relation as
indicative of a correlation
%between       $ k_\perp$ and $k_z$,
between the perpendicular and parallel sizes of
turbulent eddies.
Such eddies will be
anisotropic and elongated
in the $\vB_0$ direction.
As
        $ k_z \propto k_\perp^{2/3} $
the implied  anisotropy becomes more pronounced at smaller scales.

Finally, GS95 posit that the modal energy spectrum---for IR
scales---can be obtained by
multiplying the crude estimate for the $\vk$-space modal energy density,
namely
        $ | \delta v(k) |^2 / (k_\perp^2 k_z ) $,
by a shaping function $f$,
whose role is to strongly attenuate this estimate
        away from
the equal timescale curve:
\begin{equation}
      \Emod ( \vk )
        \sim
      \frac{ \delta v_L^2} {k_\perp^{10/3} L^{1/3}}
        \;
        f \left( \frac{k_z L^{1/3}} {k_\perp^{2/3}}
          \right)
    ,
  \label{eq:GS95spectrum}
\end{equation}
where the argument of $f$
is an approximation to
        $ \tauNL^\perp / \tauA $
  [see discussion around Eq.~(\ref{eq:tau-ratio})].
This is equivalent to Eq.~(7) in GS95,
but for physical clarity we have again used
        $ B_0 \approx \delta v_L $
to replace      $ B_0^2 $
with            $ \delta v_L^2  \approx (\epsilon_\text{inj} L)^{1/3}$;
 %% in the first numerator;
see discussion below
        Eq.~(\ref{delta-v}).
Since this is a model for the
        {IR}
spectrum, it is not expected to be valid for
        $ k_\perp \to 0 $;
its validity for
        $ k_z \to 0 $
is considered in the Commentary section below.

As introduced in GS95,
        $ f( u) $
has several properties.
It is a positive symmetric function of $u$ that is negligibly small
for
        $ |u| \gg 1 $
  (the weak turbulence modes),
and
it satisfies $ f(u) \le 1 $
and       $ \int_{-\infty}^{\infty} f(u) \, \d{u} \approx 1 $.
 %% Further features of $f$ are considered below.

Using these properties of $f$,
GS95 integrate
the modal spectrum over $k_z$ and the azimuthal
angle (cylindrical polar coordinates) to obtain the IR scaling
for the one-dimensional perpendicular spectrum, namely
\begin{equation}
        E^\perp ( k_\perp) \sim k_\perp^{-5/3}.
\end{equation}
GS95 note that this is of the same form as the Kolmogorov spectrum for
Navier--Stokes turbulence.  Fundamentally, this occurs because the
cascade timescale is the same as the nonlinear timescale.

The parallel spectrum was not determined in GS95 but turns out to be
steeper,
        $ \sim k_z^{-2} $;
see
        Eq.~(\ref{eq:Epar}).
%----------------------------------------------------

     \subsubsection{Commentary.}
                \label{sec:strong-comm}

In the previous section,
we summarized the GS95 picture of \emph{strong} turbulence,
while avoiding critical questions and commentary.
In this subsection we collect a number of these discussion points,
and in particular inquire further concerning
\begin{itemize} \itemsep=0ex
  \item the relevance of purely toroidal fluctuations;
  \item whether 3-mode couplings are present;
  \item the argument and shape of $ f(u) $,
        and the validity of the GS95 modal spectrum %%,
                                %%Equation~\ref{eq:GS95spectrum},
        for $k_z \approx 0 $;
  \item the parallel spectrum;
  \item how the cascade maintains CB at smaller scales;
  \item fluctuations with $ \tauNL (k)  \ll  \tauA (\vk) $.
\end{itemize}
These are considered in turn below.

   \paragraph{Restriction to toroidal fluctuations.}
For the GS95 strong turbulence case
with large-amplitude fluctuations and near-isotropy at the outer scale,
one may call into question the
legitimacy of the toroidal representation adopted in CB theory.
We recall at this point
that the Alfv\'en mode at
        \emph{large}
amplitude is no longer purely transverse
to the mean magnetic field $\vB_0$,
as it is for the small-amplitude Alfv\'en eigenmode.
Rather,
non-planar solutions for $\vv$ and $\vb$ fluctuations can be
found with the  {toroidal}   polarization requirement replaced by the
condition that the total magnetic field magnitude
        $ |\vB| = |\vB_0 + \vb| $
is spatially uniform\footnote{The velocity
     fluctuations $\vv$
     remain incompressible, i.e., solenoidal.}
        \citep{GoldsteinEA-sw3,Barnes76,Barnes79a,Barnes81}.
A particular class of such solutions
are polarized on the surface of a
   $ |\vB| = \text{const.}$
sphere
        \citep{Barnes81}.
Such fluctuations are routinely observed in the solar wind
at MHD scales
        \citep[e.g.,][]{MatteiniEA14,TsurutaniEA18}
and, at least in some periods,
are found to be nearly incompressible (small density variations)
and Alfv\'enic (correlated velocity and magnetic fluctuations).
So while one cannot in general rule out some admixture
of compressional turbulence, it seems to be typically small.
However, it is apparent that these
        $ | \vB | = \text{const.} $
fluctuations cannot be
represented in the basis adopted by GS95, in which
every allowed degree of freedom
satisfies       $ \vb \cdot \vB_0 = 0 = \vv \cdot \vB_0 $.

\remarkHide{
   \citep{RileyEA96} Ulysses ecliptic observations of arc-polarized $\vB$.
   Explanations:
     \cite{VasquezHollweg96,VasquezHollweg98a,
           TeneraniVelli17,TeneraniVelli18,TeneraniVelli19,
           SquireEA20}
 }

Thus the GS95 assumption of (toroidal) Alfv\'en modes
is consistent only when the amplitude is small,
contrary to their assumption of large amplitude fluctuations
at the outer scale.
Specifically, in
   the GS95
EDQNMA derivation of the CB spectrum,
only toroidally polarized fluctuations are retained
and these are not assumed to be of small amplitude.
They are also not inherently assumed to be
Alfv\'en waves
   (e.g., prescribed correlations of magnetic and velocity fields are
   not required).
The scenario adopted by the RMHD model
       (see discussion in \S\ref{sec:rmhd})
is restricted to the same toroidal
representation (Alfv\'en mode)
but also mandates  that     $ \delta b  \ll  B_0$,
a requirement explicitly absent in the `strong turbulence' version of GS95.
Furthermore,
the attempts at justifying the toroidal linear Alfv\'en mode
representation in GS95 are based almost entirely on
estimates of damping of \emph{other linear modes}
        \citep[e.g.,][]{Barnes66}
and on estimates of mode conversion
from Alfv\'en to magnetosonic modes, the argument
again grounded in linear theory.

One must conclude then, that the choice by GS95 to represent
        \emph{large} amplitude turbulence
in terms of the
        small amplitude linear Alfv\'en eigenmodes
can be questioned.
In fact, GS95 acknowledged the shortcoming of this approach to
justifying the representation,
when they stated
        (their \S5.3)
that ``\emph{\ldots our restriction to shear Alfv\'en waves
                     is no more than a guess.}''

  \paragraph{Presence of 3-mode couplings.}
In GS95 the energy-containing scales are declared to be roughly
isotropic, which implies that
        {2D modes}
are excited,
or at least are not on average
diminished relative to excitations having wavevectors
in any other arbitrary direction.
Nonetheless, one might still attempt to interpret
their strong turbulence model as lacking 2D modes, since
the GS95 development of CB
is founded on the 4-wave weak turbulence couplings.
 % In that case, the 3-mode
 %         \cite{ShebalinEA83} type couplings
 % will be absent,
 % as is also true for their `weak to strong' scenario.
But, as discussed in
        \S\ref{sec:weaktostrong-comm},
for many common boundary conditions,
4-wave interactions excite 2D modes
   \citep{NgBhattacharjee96,%
        VasquezHollweg04-jgr,VasquezEA04,HowesNielson13}.
Moreover, in simulations of MHD turbulence,
compressible or incompressible,
that are initialized with spectra lacking 2D modes,
it is typical to find significant excitation of 2D modes
within a nonlinear time
        (see \S\ref{sec:weaktostrong-comm}).
Such dynamical population (and replenishment) of 2D modes means they
%% Thus, 2D modes are populated (and replenished) dynamically and
are likely to be present in many MHD systems with a mean field.
Model spectra should reflect this, of course.

These considerations
place physical constraints on the GS95
model spectrum
     Eq.~(\ref{eq:GS95spectrum}),
and on the shape function       $ f(u) $
in particular.
For example, if $f(0) = 0$, the 2D modes are zeroed out.
This prompts discussion of these and other issues related to $f(u)$.

      \paragraph{Features connected to $f(u)$.}
As introduced in GS95, the role of $f(u)$
is apparently to localize the IR spectrum around the equal timescale
curve, and so its argument  $ u$
must obviously depend on the nonlinear and wave timescales in the IR.
A simple choice is their ratio,\footnote{Note
        that $u$ is essentially the reciprocal of the
           \emph{nonlinearity parameter},
                $\zeta_\lambda$,
        defined by Eq.~(2) in GS95.}
\begin{equation}
     u(\vk)
        =  \frac{ \tauNL^\perp( k) } { \tauA( \vk) }
        =
           \frac{ |k_z B_0|}
                {k_\perp \delta v(k) }
        \; \to \;
           \frac{ |k_z| L^{1/3}} {k_\perp^{2/3}}
  .
  \label{eq:tau-ratio}
\end{equation}
Clearly, one will have
        $ u \approx 1 $
in regions where CB holds,
and presumably also $f(u) \approx 1$ in these regions.
Here the phenomenological (right-most) approximation for $u$
is obtained after using
        Eq.~(\ref{delta-v})
and the GS95 strong turbulence requirement
        $ \delta v_L  \sim  B_0 $.
As it depends only on the components of $\vk$ this
yields an explicit form for
        $ \Emod( \vk ) $.
Taking a more self-consistent
%% (and less phenomenological)
approach, one could instead require that
        $ \delta v(k)  \approx  \sqrt{ k \Eomni( k)} $
was itself determined from the spectrum.
While
        %% perhaps
conceptually more satisfying,
this has the disadvantage of making
        Eq.~(\ref{eq:GS95spectrum})
an implicit equation for
        $ \Emod $.

What about the magnitude of $ f(u) $ when $u$ is small
  (i.e.,  $ \tauNL^\perp $   is fast)?
Even after many readings,
it is not clear to us what GS95 were intending
regarding $f$ in these
circumstances,\footnote{In subsequent works some authors have
        employed        $ f(u) \approx 1 $
        for             $   u \lesssim  1 $
        \citep[e.g.,][]{MaronGoldreich01,ChoEA02}.}
including the validity of
        $ \Emod( \vk) $
as      $ k_z \to 0 $.
There are two wavevector categories associated with
        $ u \approx 0 $:
those with
        $ k_z \approx 0 $
and those with    $ k_\perp $        very large.
The first case is obviously picking out the 2D (or nearly 2D) modes.
If it was desired to exclude these modes, one would need to have
        $ f(u) \approx 0 $
for     $ u \approx 0 $.
Possibly this was the idea in the GS95 strong turbulence model, since
their development of CB is based on the 4-wave weak turbulence
couplings and a lack of 2D excitation.  On the other hand, GS95 state
that the energy-containing scales are roughly isotropic, which
might be interpreted as indicating that 2D modes are excited.
As discussed in
        \S\ref{sec:weaktostrong-comm},
even when 2D modes are initially unexcited, they will be generated
(and replenished) dynamically.  We therefore suggest that  $ f(0) $
should be nonzero
 [and indeed that  $ f(0) \approx  f(1) $],
so that the presence of 2D modes is supported.

What does this entail for the other class of
        $ u \approx 0 $
fluctuations,
those with very large   $ k_\perp $?
Let us consider all the wavevectors in a particular
        $ k_z $
plane.
Those associated with
        $ u \approx 0 $
will have much larger
        $ k_\perp $'s
than their equal timescale
        $ u \approx 1 $
siblings.
If
        $ f(0) \approx  f(1) $
then the energy at those $\vk$-space positions will essentially only
scale with $ k_\perp $;
see
        Eq.~(\ref{eq:GS95spectrum}).
However, if
        $ f(0) \approx 0 $,
then this will cause additional attenuation of the spectral amplitude
at these large $k_\perp$'s,
acting to further localize the spectrum with respect to the $k_\perp$
directions and enhancing any ridge-like aspects of the spectrum.

Appropriate functional forms to use for $f$
are still being investigated.
In fits to
  %% moderate resolution
(driven)
numerical simulation data,
        \cite{ChoEA02}
found that
an exponential form,
        $ f(u) = \mathrm{e}^{-u} $,
gave the best agreement,\footnote{The fit
        to a Castaing function, which has the advantage of being
        differentiable near $k_z = 0 $, was also good.}
compared to Gaussian and step function options.
Note that all of these suggested forms have $f(0) \approx 1 $,
and this has important consequences.
First,
it means that $f$ does not zero out the  $ k_z = 0 $
plane and so the modal spectrum includes 2D excitation,
in general.
Second,
rather than being a ridge centred around the equal timescale
curve, the IR spectrum is more like a tilted shelf inside the
equal timescale curve, that falls off smoothly outside that curve
        \citep[cf.\ ][]{GhoshParashar15-i, ChhiberEA20-tauNL}.
Of course, as the simulations on which these fits are based are
only of modest resolution, one should be cautious about extrapolating
the results to genuinely high Reynolds number systems.

To examine
some numerical evidence that
demonstrates the above points,
we show a spectrum obtained from a free-decay $1024^3$
spectral method simulation of incompressible MHD started
from a state consistent with the GS95 assumptions.
In particular, the turbulence is strong ($\delta b / B_0 = 1$)
and the initial fluctuations are purely toroidal.
See Appendix~\ref{app:sims}
for more details regarding the code and run parameters.

\begin{figure} %[tbp]
 \begin{center}
    \includegraphics[width=\columnwidth]{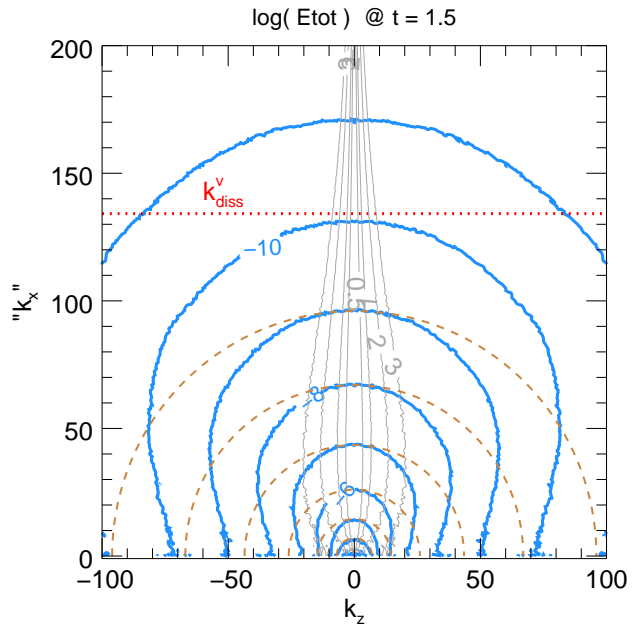}
%..    Other options:
%         height= draft, bb = a b c d, scale=0.75,
%         clip=true/false (to size of bounding box)
%         trim = 1 2 3 4 (trim (from sides): 1bp at left, 2bp at bot, 3bp at right,...)
%--------------------------------------------------------------------------------
 \end{center}
  \caption{Contour plot for a $k_y=0$ cross-section of the total
          (kinetic plus magnetic)
          modal energy spectrum,
                $ E(k_x, k_y, k_z) $.
          Data is from an
                incompressible $1024^3$
          MHD simulation, at a time shortly after that at which the
          maximum dissipation rate occurs.
          If the energy distribution was isotropic, the solid (blue)
          energy contours (solid blue) would lie on top of the dashed
          circles (brown).
          Also shown are contours for the timescale ratio,
                $ u(\vk) = \tauNL(k) / \tauA(k_z) $,
          computed using the simulation data (solid grey) and labelled with
          the value of $ \tauNL / \tauA $.
          Excitation is essentially isotropic when the timescale ratio
          is less than about 1.5.
          The perpendicular dissipation wavenumber is indicated in dotted red.
          Initial conditions for $\vv$ and $\vb$ were
          toroidally polarized fluctuations in the wavenumber band
                $ 3 \le |\vk| \le 7 $,
          with no net cross helicity, and a strong turbulence energy
          partitioning:
                $\delta v = \delta b = B_0 = 1 $,
          all consistent with the GS95 assumptions.
          Initial Reynolds numbers are $\approx 500 $.}
  \label{fig-MHDsimspectrum}
\end{figure}

    Figure~\ref{fig-MHDsimspectrum}
displays a cross-section of the computed
modal energy spectrum after
approximately one nonlinear time.
One sees immediately
that there is no indication
of a deficiency in power at either very low $ k_z $
or at   $ k_z = 0 $.\footnote{\cite{GhoshParashar15-i} find similar
  behavior for a wide range of initial conditions in compressible MHD.}
In fact the contours of spectral energy density
are almost circular within the equal
timescale curve.\footnote{Very similar figures are obtained if the
        toroidal energy is used in place of the total energy.
        Indeed,
        in both the incompressible and the compressible MHD situations,
        simulations started with fluctuations having either
        isotropic variance or toroidal variance,
        yield similar $ k_\perp $--$ k_z$  energy contour plots.}
This is consistent with the idea that
spectral transfer in this region should become progressively
more isotropic as one moves deeper into the region
in which the nonlinear rate is dominant
  (cf.\ discussion later in this section
   and in \S\ref{sec:strong-frustrated-iso}).
Assuming that the IR spectrum can be described with a CB-like model,
the numerical results also indicate that
        $ f(u) \approx f(1) $
when    $   u \lesssim 1 $,
and in particular that
        $ f(0) \ne 0 $.
Note also that there is no sharp change in the spectrum as the
timescale ratio contour with
        $ u(\vk) = 1 $ (or similar values)
is crossed
        \citep[e.g.,][]{VerdiniGrappin12},
indicating that $ f(u) $ should change relatively slowly for $u$ near 1.
In particular, a step function is quite a drastic simplification for
the the form of $f(u)$.
%-----------------------------------------------------------------------------%
%-----------------------------------------------------------------------------%

        \paragraph{Parallel spectrum.}
Although not calculated in GS95, it is
straightforward to obtain this reduced spectrum\footnote{Recall
        the classical definition of a reduced spectrum is one obtained
        by integrating over all but one of the $\vk$-space
                \emph{Cartesian}  coordinates
        \citep{BatchelorTHT}.}
from the modal IR spectrum they present,
stated herein as
        Eq.~(\ref{eq:GS95spectrum}).
Assuming axisymmetry and integrating over the two
        $ \vk_\perp$ coordinates
leads to
\begin{equation}
      \Epar( k_z)
        =
      2 \pi \int_0^\infty k_\perp \Emod(\vk) \, \d{k_\perp}
        \propto
      \frac{1} {k_z^2} .
  \label{eq:Epar}
\end{equation}
Rather remarkably this
        $ k_z^{-2} $
scaling is independent of the functional form of
        $f$,
provided that, in addition to the properties listed in GS95, it also
has a finite first moment:
        $ \int_0^\infty u f(u) \, \d{u} < \infty $
        \citep{ChoEA02}.\footnote{A step function
    approximation to $ f(u) $,
    with, $ f = 0 $ for $u > 1$, say,
    makes it particularly clear that the $ k_z^{-2} $
    scaling is a consequence of excitation located \emph{inside} the
    equal timescale curve, since there is then
        \emph{zero}
  excitation outside the curve.}
Thus
        $ f(u) $
needs to fall off faster than
        $ 1/u^2 $
as      $ u  \to  \infty $.
Nonetheless,
one must be just a little wary of the
        $ k_z^{-2} $
scaling as
        Eq.~(\ref{eq:GS95spectrum})
was not developed with validity near
        $ k_\perp = 0 $   ($ \Rightarrow u \to \infty $)
in mind, but these values have been integrated over.

Provided there really is negligible energy associated with
        $ k_\perp \approx 0 $
modes, the above parallel spectrum scaling result should be correct for
        $ 1  \ll  k_z L
              <  (\delta v_L / B_0)  (k_\perp^\text{diss} L )^{2/3} $.
At still larger
        $ k_z $,
the (steeper) perpendicular dissipation
range is encountered before the equal timescale region.
Because of this, the integration over
        $ \vk_\perp $
does not pick up sufficient energy to give the
        $ k_z^{-2} $
scaling,
since it does not traverse a  $\vk$-space region
of substantial enough excitation.
See
        Figure~\ref{fig-CBweak}(b).
This implied
cutoff in the $k_z^{-2}$ parallel spectrum,
based on encountering the dissipation range in $k_\perp$,
is discussed extensively in
    \cite{TesseinEA09}.

Solar wind observational studies can determine one-dimensional
(reduced) spectra, or wavelet analogs of them,
as a function of the angle between the mean magnetic field and the
wind speed (observation) direction,
        $ \theta_{U\!B} $.
Some of these studies provide support for a CB spectrum,
finding a smooth transition from a spectral slope of
        $ \approx -5/3$
at large
        $ \theta_{U\!B} $
to      $ \approx -2$
at small, nearly parallel, angles
        \citep[e.g.,][]{HorburyEA08,Podesta09-SWaniso,DuanEA18}.
However, other studies do not,
finding the same slope for all
        $ \theta_{U\!B} $,
within errors
           \citep{TesseinEA09,WangEA16-notCB,TelloniEA19-notCB,WuEA20-notCB}.
See
        \S\ref{sec:observations}
for further details.

Note that other arguments for the form of the parallel spectrum
exist.
For example,
        \cite{Beresnyak15}
has proposed
that the parallel spectrum calculated along (local) field lines
is really the Lagrangian frequency spectrum in disguise and does not
involve any use of CB.
Intriguingly, this leads to
the same $ k_z^{-2} $
scaling associated with CB.

   \remarkHide{The CB form is
                $ \Epar( k_z) = \text{const.}
                            \frac{ \epsilon^{2/3}} { L^{1/3}}
                             k_z^{-2} $,
        with no explicit $B_0$ dependence
          (more or less in the usual Kolmogorov way). \newline
        The Lagrangian-frequency-spectrum-in-disguise form (also
        given by the RMHD `preserve Alfv\'en symmetry' argument), however,
        has quite different dependence on $\epsilon$ and $B_0$:
                $ \text{const.} \frac{ \epsilon} {V_A} k_\parallel^{-2}$.
      }

\paragraph{Why does the cascade preserve/propagate
              the CB property, once it is established?}
This is a question relevant to both the `weak to strong'
and the `initially strong' CB scenarios.
The CB wavenumber relation,
        $ k_z  \sim  k_\perp^{2/3} $,
may be viewed as a consequence of this feature of the cascade and
indicates that
both perpendicular and parallel transfer are active in a critical
balance state.
In the IR one can argue as follows.
Velocity and magnetic perturbations at a scale
        $ \ell  \ll  L $
are small,
        e.g.,  $ \delta b(\vk)  \ll  B_0 $,
where
        $ | \vk |  =  1 / \ell $.
Although these fluctuations have small amplitudes
their nonlinear effects
may be either weak or strong
        (relative to the linear effects)---depending
upon the
        \emph{orientations}
of their wavevectors
        (Figure~\ref{fig-weakVstrong}).
How energy is distributed over weak and strong fluctuations at
scale   $ \ell $
obviously depends on the nature of the energy cascade.
Because
fluctuations near (or inside) the equal timescale curve,
are only
        weakly
aware of the wave timescale,
they will engage in spectral transfer that is not so different
from the isotropic Kolmogorov cascade,
moving excitation to
 %%        ``destination''
wavevectors on shells of somewhat
larger $\vk$-space radius.
 %%        $ k_\text{new} $, say.
Crudely, we may divide the $\vk$'s on these ``destination shells''
into three categories: those that are either
(i) well inside the $\tauNL \approx \tauA $ zone,
(ii) in or near that zone,
or
(iii) well outside it (i.e., at larger $k_z$).
See
        Figures~\ref{fig-weakVstrong}
        and~\ref{fig-CBweak}.
Naturally, when transfer is to elsewhere in the
equal timescale zone, this assists with continuation of CB to
smaller scales.

What about transfer in the third category?
This can involve ``destination''
        $ k_z $s
that are large enough to have
        $ \tauA(\vk)  \ll  \tauNL(k) $;
that is, these modes will be weakly turbulent.
As already discussed,
their dominant dynamics is perpendicular spectral transfer,
with a statistical tendency towards higher $ k_\perp$.
For
        $ k_z$
not too large this perpendicular transfer is also back towards the
equal timescale zone
        (Figure~\ref{fig-CBweak}).
Once the excitation arrives there, the physics changes again, this
time back to
(frustrated) isotropic spectral transfer.
Thus, one might be tempted to describe the equal timescale zone as
being
        `attracting.'
This, however, is inaccurate since the perpendicular transfer process
does not involve any seeking of the zone, or even awareness of it
  (recall that the process is mediated by 2D modes).
Section~\ref{sec:attract}
discusses this characterization as `attracting' in more detail.

Summarizing, because wave-like effects
are not dominant near the curve, modes in this region experience
roughly isotropic spectral transfer.
However, some of that transfer places excitation at
        $\vk$-space positions
where $\tauA$ is the faster timescale,
bringing weak turbulence effects into play there---most notably,
strong perpendicular transfer and movement of excitation back towards
the     $ \tauNL  \approx  \tauA $
zone.
%% there is a tendency for excitation that moves outside
%% the equal timescale zone to be shepherded back towards it.
%% (at a higher $k_\perp$ part of) the zone.

   \paragraph{Strongly turbulent fluctuations.}
We have not yet considered the first category of destination
$\vk$'s---involving transfer from inside the curve to further inside
the curve, where
        $ \tauNL( \vk) $
is considerably smaller than
        $ \tauA(k_z) $.
Section~\ref{sec:corrn-length}
discusses the evolution of these
  strongly turbulent fluctuations
and the roles they play in the dynamics.
It is worth emphasizing that in general such modes
can be
present in MHD turbulence.
This has sometimes not been adequately appreciated,
perhaps due to blurring of the distinction between
(i) rms fluctuation amplitudes at some scale
and
(ii) the correlations associated with that scale.
Obviously the former are only small if almost all the individual
contributions are small,
whereas some correlations can be zero even when the
fluctuations are large-amplitude.
In particular, a lack of correlation at some scales certainly does not
imply a lack of excitation at those scales. When large-scale
fluctuations are uncorrelated, one expects the corresponding
range of small wavenumbers to exhibit a flat spectrum at a nonzero
power level
        (cf.\ \S\ref{sec:corrn-length}).

This completes our commentary on the GS95 presentation of CB.
In the remainder of the paper we take up various related issues
in greater detail.

%===============================================================

        \section{Local Mean Field vs. Global Mean Field}
        \label{sec:local-field}

  The original GS95 presentation was based on
considerations of perpendicular and parallel with respect to the
        \emph{global}  mean field, $\vB_0$.
This is implicit in the phenomenology they present
and explicit in their EDQNMA closure calculations.
The 1997 ``intermediate turbulence'' model
of Goldreich and Sridhar
        (GS97)
makes no substantive use of the local field beyond a single
passing allusion to the arguments of
        \cite{MontMatt95} and
        \cite{NgBhattacharjee96}
which, contrary to the GS97 suggestion,
     \remarkHide{end of their first para, p680}
actually both employ
        \emph{uniform global}
mean magnetic fields in their analyses,
and not local mean fields.\footnote{We remind the reader
    that we
    employ the notation $ k_z$ for wavenumber parallel
    to the global mean field $\vB_0$,
    and $ k_\parallel$ for the wavenumber component
    parallel to a local mean of $\vB$.}

To the best of our knowledge
the first paper to articulate a need to employ a local
mean field direction in reference to the
development of spectral anisotropy is
        \citet{ChoVishniac00-aniso}.
A nearly contemporaneous paper employed
a different analysis method     \citep{MilanoEA01},
and
also concluded that conditional
second-order structure functions show a greater
degree of correlation anisotropy when measured relative
to a locally computed mean field direction. Interestingly, this
result also obtains when there is no  uniform DC component
of the magnetic field,
so that the global spectrum is isotropic.
Based largely on the local mean field formulation
in      \S5
of         \citet{ChoVishniac00-aniso},
a popular reinterpretation of the CB of GS95 emerged, as
stated in \S6.6 of \citep{MaronGoldreich01};
namely that the proper CB approach should be based on a
        \emph{local}
mean magnetic field.

\remarkHide{ChoVishniac00-aniso use 2 methods to calculate
      $ \avg{\vB}_\text{local} $.
    1. 2-point average.
    2. Gaussian envelope/convolution.  But this is NOT used to
      examine scalings. }

Returning to the apparent source,
       \citet{ChoVishniac00-aniso}
showed that if the lengthscale anisotropy increases with decreasing
scale then a Fourier transform analysis can mask the actual scaling
        (say $ k_z \propto k_\perp^{2/3} $),
yielding a
        $ k_z \propto k_\perp $
scaling instead.
Note that the latter linear scaling
may be related to
scaling with energy-containing range timescales \citep{OughtonEA98}.
It should be emphasized that
straightforward global mean field analysis had already found that
perpendicular spectral anisotropy increases at smaller scales,
based on MHD simulation results in
both 2D \citep{ShebalinEA83}
and 3D  \citep{OughtonEA94}.
Therefore one may firmly conclude that the
        \emph{existence}
of scale-dependent perpendicular anisotropy does \emph{not} depend on
reference to a local mean field direction.
It is equally clear
that perpendicular anisotropy measures are indeed greater when
measured relative to locally determined mean magnetic fields
        \cite[e.g.,][]{ChoVishniac00-aniso,MilanoEA01,MattEA12-local};
that is, the
        \emph{magnitude}
of the scale-dependent anisotropy does vary
with the choice of local mean field calculation.

For example,
when estimates for
        $ \ell_\parallel $
and     $ \ell_\perp $
were calculated relative
to a two-point approximation for the local mean field, results from
incompressible 3D MHD simulations with
        $ \delta b / B_0 \approx 1 $,
have frequently found consistency with the
        $ \ell_\parallel \propto \ell_\perp^{2/3} $
GS95 scaling
      \citep[e.g.,][]{ChoVishniac00-aniso,ChoEA02}.
In addition,
numerous solar wind observational studies
(usually employing the magnetic field data)
have reported similar agreement with the GS95 spectral scalings when
the analysis is performed with respect to a local mean magnetic field
        (see \S\ref{sec:observations}).

We should note,
on the other hand, that
there are solar wind observational studies
that
do
        \emph{not}
support the presence of CB scalings.
Thus the available evidence
makes it difficult to draw firm conclusions on this
     (see also \S\ref{sec:observations}).
For example,
   \citet{WangEA16-notCB}
use a wavelets analysis to show that the 5/3 to 2 slope transition as
        $\theta_{U\!B} \to 0$
is      \emph{not} really seen
when one demands more stationarity of the local mean field.
   (Here
        $ \theta_{U\!B} $
   is the angle between the mean magnetic field and
   the wind (observation) direction.)
In particular, if
        $ W( \tau_m, t_k) $
is the wavelet coefficient at time $t_k$ and timescale $\tau_m$, they
only allow this to contribute to a particular $\theta_{U\!B}$ bin
  (say $0^\circ$--$10^\circ$),
if
        $ \theta_{U\!B}(t) $
is in that bin at (at least) the three times
        $ t_k \pm 1.5 \tau_m $ and $t_k$.
These times correspond to approximately the start, end, and middle of
the interval used to  calculate the wavelet coefficient at that scale.
  (Other wavelet studies have typically only imposed a requirement on
        $ \theta_{U\!B} $
   at the single time $ t = t_k $
       \citep[e.g.,][]{HorburyEA08,Podesta09-SWaniso}.)

In an earlier study
        \citet{TesseinEA09}
employed a traditional
        (Fourier)       %%(non-wavelet)
analysis approach and mean fields
determined from the whole interval, obtaining results very similar to
those of
        \cite{WangEA16-notCB}.
More recently
        \cite{TelloniEA19-notCB}
used Hilbert spectral analysis
        %% (empirical mode decomposition)
on solar wind data and reported that magnetic power spectra in the
field-aligned direction exhibit
        $ k_\parallel^{-5/3} $
scaling, rather than the $ k_\parallel^{-2} $ expected for CB.
\changed{Similarly,
    in an analysis of
        \emph{Wind} data
    with strong requirements
    on the directional stability of the local mean field,
        \cite{WuEA20-notCB}
    showed that the parallel and perpendicular scaling exponents are
    essentially the same for the second-order structure functions:
    $ \approx 2/3 $ for the magnetic fluctuations
    and $ \approx 1/2 $ for the velocity ones.
}

One concludes, then, that properties
obtained only using local mean field estimates
may be stable only when certain conditions are attained
  (e.g., see
        \cite{Panchev,GerickEA17,Podesta17-rms,IsaacsEA15}).

        %.. Panchev \S13, p74-78. The influence of finiteness on the
        %interval of averaging.

This sensitivity to regional conditions
leads us to an important fundamental contrast between global correlation
analysis of anisotropy and the local mean field version.
When analysis is based on a
        \emph{local}
mean field
  (e.g., using wavelets or
        structure functions relative to a locally computed mean),
the statistical order of the calculated moment is higher than
a naive identification would suggest.
In particular, the order is higher than
the equivalent moment calculated relative to a globally defined
mean field
        \citep[e.g.,][]{MattEA12-local}.
For example the structure function
        $ \avg{ | \vb(\vx) - \vb(\vx + \ell \ehat) |^2} $
is of second order provided
        $ \ehat $
is a fixed direction.
If, however, one associates $ \ehat $
with the direction of a local mean field,
this will vary with position and     $ \ehat(\vx)$
is itself a random variable.
For such approaches, the local mean-field structure function is, in
general, of higher-order (than second).
This property is related to the fact that the local mean field
analysis may be viewed as a conditional statistic.
 %% See
 %%        \S\ref{sec:local-field}.
Clearly, this characterization
also applies to observational and simulation
studies, some of which are discussed in
        \S\ref{sec:observations}.

The scale-dependent nature of perpendicular anisotropy is firmly
established.  So, too, is the property that anisotropy is greater
relative to a mean field that is locally calculated.
The remaining questions regard
the physical properties of what one calls a ``spectrum''.
As just discussed, the extra condition imposed by
a wavelet decomposition or a structure function
that depends on local mean field values
introduces an additional random variable.
The resulting statistic is no longer merely a second-order moment of
the joint distribution of the magnetic field components.
Instead, it is a higher-order moment, in effect a conditional statistic,
that no longer satisfies the familiar property that spectra are
insensitive to phase randomization.
It follows that physical properties
  (such as enhanced perpendicular anisotropy)
that depend on the local mean field are related to non-Gaussian
statistics and therefore related to intermittency
        \citep{Novikov71,SreenivasanAntonia97}.
Accordingly, phase randomization destroys the dynamically produced
enhanced perpendicular anisotropy relative to the local magnetic field,
as can be straightforwardly demonstrated using MHD simulation data
        \citep{MattEA12-local}.

Note that in RMHD (see \S\ref{sec:rmhd})
the distinction between local and global mean fields is immaterial.
This is because the RMHD model is based on the presence of a strong
mean field such that
        $ \delta b, \delta v \ll B_0 $.
Thus the local mean field is, to very good approximation, the same
as the global mean field in the RMHD limit.

%===============================================================

%% \clearpage
        \section{RMHD: Contrast with CB Derivation}
                \label{sec:rmhd}

Various similarities connect
CB approaches and
the Reduced MHD approximation.
 %%        \citep{KadomtsevPogutse74,Strauss76,%
 %%                Mont82-strauss,ZankMatt92-rmhd,OughtonEA17-rmhd}
The most important of these is that they are both models for strong
anisotropic turbulence.
Hence, energetically speaking, any effects associated with linear waves
are secondary,
or perhaps comparable,
to the nonlinear activity.
Because of this, the equal timescale curve,
        $ \tauNL( \vk) \approx \tauA( k_z) $,
features prominently in each model.
It does so, however, in distinct ways and we now discuss these
differences.

Recall that
CB phenomenologies are typically developed for IR fluctuations and
posit that the most important ($\vk$-space)
dynamics is associated with wavevector modes on or near
the equal timescale curve.
This is sometimes stated as CB holds scale by scale in
    the IR
	\citep[e.g.,][]{NazarenkoSchekochihin11}.
One outcome is usually a
functional form for the IR energy spectrum.
In this section we are drawing comparisons with the RMHD
approximation, so
we focus on the original system to which CB arguments were applied:
incompressible MHD with a mean magnetic field of
        \emph{moderate}
strength,       $ \delta b /B_0 \approx 1 $
        (GS95).
In that work the poloidal
        (pseudo-Alfv\'en mode)
fluctuations are simply discarded at the beginning,
with arguments for their neglect being advanced later
        (cf.\ \S\ref{sec:otherFlucts} below).

The starting state for deriving RMHD is quite different.
One begins with compressible MHD threaded by a
        \emph{strong}
mean field.
The fluctuations are thus energetically weak
        (i.e., of small amplitude: $ \delta v, \delta b  \ll B_0 $)
and one might imagine that the leading-order behavior is associated
with linearized waves.
However, one can instead ask, under what conditions
can these weak fluctuations
have a leading-order dynamics that is nonlinear.
The key requirement
turns out to be the elimination of all high-frequency fluctuations,
meaning those with timescales faster than the nonlinear one
        \citep{KadomtsevPogutse74,Strauss76,Mont82-strauss,
                ZankMatt92-rmhd,
                SchekochihinEA09,
                OughtonEA17-rmhd}.
Following this procedure yields the RMHD model provided
some (leading-order) restrictions are imposed:
\begin{enumerate}
  \item[1)]
     Spectral anisotropy is present with $ k_\perp \gg k_z $.
        This ensures the absence of high-frequency Alfv\'en modes
        and high-frequency slow modes.
  \item[2)]
     Parallel variances for $\vv$, $\vb$ are zero: $ v_z = 0 = b_z$.
     This ensures the absence of high-frequency fast modes.
\end{enumerate}
Although not an objective, these conditions in fact lead to the
elimination of \emph{all} slow and fast modes,
not just the high-frequency ones.
Consequently the only fluctuations that remain are incompressible
        toroidal  ones,
and---\emph{by construction}---these have
\begin{equation}
     \tauNL( |\vk|)  \lesssim  \tauA( k_z ),
 \label{eq:RMHDcondit}
\end{equation}
which one may call the
        \emph{RMHD timescale condition.}
We note also that
from standard
derivations of RMHD in a uniform mean magnetic field,
the emergent dynamics are incompressible,
with no density fluctuations.
Along with the vanishing of the parallel variances,
this amounts to a representation that is
structurally similar to what is called
the Alfv\'en mode in the
small amplitude limit. This is also the representation
adopted in GS95 as a basis for CB
(but without RMHD's requirement that $B_0$ be large).

It is also apparent that Eq.~(\ref{eq:RMHDcondit})
is
reminiscent of the CB condition,
        Eq.~(\ref{eq:CB-condit}).
However, it is not the CB condition and
the difference is important:
RMHD fluctuations have wavevectors that can lie
        \emph{anywhere}
inside the equal timescale curve, whereas the GS95 version of CB seems
to include only modes on or near the curve.
 %%       (Figure \ldots).
In particular, in RMHD modes with very small
        $ k_z $
are dynamically important.\footnote{Note that
        the RMHD model does not really support a linearized version of
        itself. If fluctuation amplitudes are so small that
                $ \tauNL > \tauA $
        then the assumptions leading to the RMHD model are no longer
        valid (except for the 2D modes).}
This includes the case of
    $ k_z = 0 $ (2D) fluctuations.\footnote{Indeed
        pure 2D turbulence has been considered
        as a subset of RMHD for some purposes,
        including attaining higher resolution
        coronal heating models
        \citep[e.g.,][]{EinaudiEA96,DmitrukEA98}.}
Clearly, the latter have no wave character associated with the strong
        $ \vB_0 $,
and this is a significant difference from the CB condition
with its assumption that the wave timescale is always relevant.
This distinction between the RMHD and CB timescale conditions has not
always been appreciated.
For example, GS95 (p.~774) remark that
    \emph{``critical balance between parallel and
         perpendicular timescales is a key assumption in the
         derivation of the Strauss equations''},
and, as we have just discussed, this statement is
too restrictive, and in fact imprecise.

In an RMHD system \emph{all} fluctuations have
	$ k_\perp \gg k_z $
so that the equal timescale curve is `close' to the
	$ k_z =0 $  plane.
Suppose that the rms fluctuation level is given as
	$ \delta v, \delta b \approx 1 $.
If these are RMHD fluctuations (with $B_0 \gg 1$)
they will occupy a      $ \vk $-space region
that is
considerably narrower and more anisotropic
relative to the region associated with a GS95 CB model
   (for which $\delta v, \delta b \sim B_0$).

The similarity between
the
RMHD requirement
        $ \tauNL(\vk) \lesssim \tauA(\vk) $
and the CB condition
       $ \tauNL(\vk) \approx \tauA(\vk) $
has sometimes
blurred the distinctions between these theories.
Some studies of RMHD systems,
or systems that reduce to RMHD,
have led to results
claimed to be demonstrations of CB,
but are arguably consequences of RMHD.
This is because
various quantities within RMHD
have scalings consistent with CB
        \citep{ChandranEA15-intermitt,
                MalletEA15,MalletEA16},
or even indistinguishable from CB.

An example serves to demonstrate this point.
Consider a spectrum of anisotropic
fluctuations in the presence of a strong $B_0$.
Let us further assume a Kolmogorov
phenomenology, including an anisotropic
perpendicular $k^{-5/3}$
spectral law \citep{FyfeEA77b,MontTurner81}
and scale locality.
This a perfect setup for RMHD to apply, provided that fluctuations
violating the RMHD condition are absent
  (i.e., there are no high frequency wave-like fluctuations).
We may then ask:
What is the condition on parallel
wavenumber $k_z$ to ensure this hydrodynamic-like behavior,
considering here, for simplicity, only the inertial range?
Equivalently,
        \emph{what is the maximum bandwidth in  $ k_z $
              of the RMHD inertial range fluctuations,
              as $k_\perp$ is varied?}
The calculation is straightforward:
Consider a perpendicular wavenumber $k_\perp$.
The spectrum is highly anisotropic
so that the populated portions of
shells of radius $k$ are almost indistinguishable
from slivers of $k_\perp$ ($\approx k$).
Let the mean-square parallel extent of the spectrum
        (a function of $k_\perp$)
be written as
   $ \langle k_z^2 \rangle $
and assume that the
parallel spectrum extends until the
RMHD timescale condition          Eq.~(\ref{eq:RMHDcondit})
is marginally violated.
Performing the calculation one finds that in the IR
the     rms  (energy-weighted)
parallel wavenumber is
\begin{equation}
  \overline{k_z}  = \langle k_z^2 \rangle^{1/2}
                = C \frac{\epsilon^{1/3}}{B_0}  k_\perp^{2/3} ,
  \label{eq:rmhd-scaling}
\end{equation}
where $C$ is a constant.
Thus, in this simple example
the maximum
parallel extent of the RMHD spectrum follows precisely
the Higdon curve        Eq.~(\ref{eq:Higdon-k}),
consistent also with the CB equal timescale curve
        Eq.~(\ref{eq:CB-condit}).

It is perhaps not surprising that some RMHD quantities scale
similarly to CB expectations.
Nonetheless,
for quantities that depend crucially on low $k_z$
(quasi-2D) activity,
the reasons for scalings to be of the CB types
are less directly motivated.
For example, the rate of 3-mode perpendicular transfer depends on the
amplitude of the (quasi-)2D fluctuations;
due to their small
values of $k_z$ and implied very long wave period,
these do not satisfy the CB timescale condition of
        $ \tauNL \approx \tauA $
        \citep{ShebalinEA83,Grappin86}.
Therefore it seems reasonable to conclude
that a realization of
MHD turbulence in which a significant
portion of the cascade is due to
these quasi-2D fluctuations is not
 ``critically balanced''
in the GS95 sense.
However such
spectral distributions that
lie     \emph{within}
the bounds expressed in
        Eq.~(\ref{eq:rmhd-scaling})
may still be described
in an RMHD representation.

At this point it is opportune to mention a
%% very interesting
paper
        \citep{MalletEA15}
that examines, within the IR,
the probability distribution of
the nonlinearity parameter      $\chi = \tauNL / \tauA $
conditioned on the scale $\lambda$,
        i.e., the probability distribution $ P( \chi | \lambda)$.
Using numerical experiments,
the authors note that the nonlinear
time and the Alfv\'en time in the IR have
non-selfsimilar distributions,
but their ratio $\chi$
has a distribution that collapses for IR lags.
This is the essence of their main result:
that the
   ``nonlinearity parameter $\chi^\pm$ has a scale-invariant
     distribution \ldots''.

The  \cite{MalletEA15}
result is indeed an interesting perspective, and in
        \S\ref{sec:observations}
below we will discuss the subtle issue as to whether
that conclusion actually provides evidence for a general appearance of
the critical balance spectrum.
Here, though,
we raise the question
as to whether investigation of the dynamical
development of anisotropy in MHD should properly be carried out
using an initial        \emph{assumption}
of an RMHD model.
Both RMHD and CB are models
intended to describe anisotropy of full MHD.
However,
from the timescale perspective
CB is a subset of RMHD,
as is readily seen by comparison of
the fundamental CB assumption
         $\tauNL \approx \tauA $
and the RMHD timescale condition
         $\tauNL \lesssim \tauA $.
On the other hand, from the perspective of $B_0$ the models can be
distinct, with RMHD derivations requiring a strong $B_0$,
while the GS95 derivation of CB imposes only a moderate mean field,
        $ \delta b \approx B_0$.
The key point is that RMHD already has built into its framework
the conditions and dynamics that
lead to spectral anisotropy and the
elimination of fast timescales.
The     \emph{development}
of this anisotropy is embedded in the derivation of the
respective models.
It therefore seems logically
flawed to assume RMHD at the onset
in examining whether
anisotropy actually emerges,
since RMHD already assumes many---but not all---of
the properties
that would be needed to establish CB.
We are therefore skeptical
of the generality of conclusions regarding CB
that emerge in strategies based on RMHD models
        \citep[e.g.,][]{MalletEA15}.

RMHD and CB (as applied to incompressible MHD)
also have some other significant
differences.
Most fundamentally, the
RMHD approximation yields a set of equations for the evolution of the
fluctuation \emph{fields},
        $ \vv (\vx, t) $        and     $ \vb (\vx, t) $.
Whereas the CB condition facilitates a phenomenological approach which
can provide approximate forms for energy \emph{spectra}.
Information about the velocity and magnetic fields  is not obtained.
Moreover, CB models usually assume an underlying steady state
of the turbulence
and thus do not include any time evolution.
This can be contrasted with considerations of evolving (R)MHD that
approach a steady state.
For example, the RMHD model,
like the full 3D MHD model,
 can accommodate a state in which 2D modes
are absent initially and become populated at a later time.
As discussed in         \S\ref{sec:weaktostrong-comm},
this scenario presents obvious problems to a CB approach.

Finally, we recall that
the RMHD model is founded on the small amplitude
approximation,
        $ \delta v, \delta b \ll B_0 $,
and therefore
the distinction between the global mean field and the local mean field
does not arise: to very good approximation, the two are equal.
If, however, one was interested in a local
mean field associated with a
transverse direction, the global and local averages
would in general differ.

%----------------------------------------------------

       \section{Wider Applicability of the CB Curve}
                \label{sec:curve}

As emphasized herein and elsewhere
        \citep[e.g.,][]{SchekochihinEA09,NazarenkoSchekochihin11}
the equal timescale curve (or more realistically, zone) is of wider
applicability than the GS95 context of incompressible Alfv\'enic MHD
turbulence with $\delta b / B_0 \approx 1$,
and indeed predates that work
    \citep[e.g.,][]{MontTurner81,Higdon84}.
In a sense the idea is elementary, in that a system with
timescales associated with two distinct processes may well have regions
        (in $\vk$-space or $\vx$-space)
where these timescales are approximately equal.
Nonetheless, the recognition of this in the context of MHD turbulence
was an important advance.

Recently, \cite{Terry18}
has noted that CB is most appropriately regarded as a hypothesis whose
basic premise is that the shortest time to spatial decorrelation
essentially sets
a single overall correlation time.  This is related to
        Kraichnan's
        (\citeyear{Kraichnan65})
insight
that the energy cascade rate should be directly proportional to the
triple correlation timescale, $\tau_3$,
associated with the lifetime of terms like
        $  \avg{\vv \cdot \left( \vv \cdot \nabla\right) \vv} $.
In general, many processes can contribute to the decorrelation of the
triples---advection,
        shear, wave propagation, dissipation, etc---but
in some situations one process may be dominant.
Some of these special cases are well known.
For example when advection dominates a
Kolmogorov      $ k^{-5/3} $
spectrum emerges,
whereas when Alfv\'en wave propagation is the primary MHD
decorrelation mechanism the Iroshnikov--Kraichnan $ k^{-3/2} $
spectrum ensues---as long as
        $ \tauA $
is approximated as      $ 1/(k B_0) $
        \citep{Iroshnikov64,Kraichnan65,MattZhou89-IR,ZhouEA04}.
CB can be viewed as a spectral form that arises when advection and
wave propagation make comparable contributions to the triple
decorrelation rate.

Clearly, the equal timescale zone plays an important role in RMHD,
serving as a rough outer boundary between those Fourier modes
that formally satisfy
the conditions needed for the RMHD approximation to hold from those
that do not.
The zone is also important in defining quasi-2D MHD fluctuations
  (see the glossary).
These have
        $ k_\perp \gg k_z $
 {and}
        $ \tauNL < \tauA $,
and thus lie inside the equal timescale curve.
Recall that in GS95 parallel components of the
        $\vv$ and $ \vb$ fluctuations are neglected.
However, MHD systems that
have parallel variances can also have
equal timescale curves.
This includes compressible MHD, although that is a more
complicated situation.  There, parallel variances can be
associated with compressive fluctuations so that there are additional
wave timescales to consider (fast and slow mode), together with the
Alfv\'en timescale.

Even systems with a weak mean field still have an equal timescale curve.
In the $B_0 \to 0 $ limit,
the curve recedes to $k_z \to \infty $
so that the     $ \tauNL < \tauA $
region covers the entirety of $\vk$-space, as is to be expected for
statistically isotropic situations
        \citep{OughtonEA06-2cpt}.

%===============================================================

        \section{Is the CB Curve Attracting?}
        \label{sec:attract}

GS95 state
      (p.~774)
that the turbulence
        \emph{``self-regulates so there is an approximate balance between''}
        $ \tauA $ and $\tauNL $.
Thus, one might readily interpret their arguments as meaning that
once excitation reaches the CB (aka equal timescale) zone,
further spectral transfer occurs within this
%%        $ \tauA(|\vk|) \approx \tauNL(k_z) $
zone.
   \cite%[p.~3]
        {GaltierEA05}
express this viewpoint strongly, stating that the CB curve
  \emph{``may be seen as a path in the
            \emph{($k_\perp$--$k_\parallel$)}
   Fourier space followed naturally by the dynamics  \ldots.
   In other words,
        {it means that excitations are concentrated on this
          curve:}
   the curve does {not} define a boundary between regions where
   wave or strong turbulence dominates.''}
\cite{Alexakis07}
has described a related phenomenology.

Our perspective on this differs
        \citep{OughtonEA04-rmhd,OughtonEA06-2cpt}.
The curve is by definition a
        %% approximate
boundary in Fourier space,
separating regions where
        $ \tauNL / \tauA < 1$
from regions where this ratio exceeds unity
        (cf.\ \S\ref{sec:tauNL}).
As this is the essential distinction between the primacy
of strong turbulence
versus the dominance of weak turbulence effects,
we also regard the curve as delineating $\vk$-space regions
of strong and weak turbulence.
Since spectral transfer processes are unaware of the
existence of a CB curve, never mind actually seeking it,
it is inaccurate to describe the curve as attracting.
The curve does, however, serve as a loose outer boundary to a region
        ($ \tauNL \lesssim \tauA $)
from which it is relatively hard for excitation to escape
(at least for long---see below).

For fluctuations outside the curve
the wave timescale is faster and
spectral transfer is predominantly perpendicular
        (Figure~\ref{fig-CBweak}b).
This process does tend to move
the excitation towards the CB curve,
provided  $k_z$ is not too large.
However, this transfer to higher
        $ k_\perp $
is not a dynamical effort to reach the CB curve---it is just
perpendicular transfer that might, or might not, encounter the curve.
If the curve is encountered, excitation is moved
into the equal timescale zone and this
means the physics changes:
at these $\vk$-space locations the wave timescale
is now slower than the nonlinear one, and spectral transfer is
approximately isotropic.
See     \S\ref{sec:strong-comm}.

As we have already discussed,
below   Eq.~(\ref{eq:Epar}),
when $k_z$ becomes sufficiently large, the associated dominant
perpendicular transfer moves energy to the perpendicular dissipation
scale without encountering the CB curve at all.  These are the
        $ E( k_\perp) \propto k_\perp^{-2} $
fluctuations of weak turbulence
        \citep{GaltierEA00}.
Estimates
of the  $ k_z $
at which this spectral transition occurs are discussed in
        \cite{TesseinEA09}.

Is there attraction from inside the curve (i.e., at low $k_z$)?
No, not in the sense that these are preferred locations for the
excitation.
To a reasonable approximation, spectral transfer is isotropic in
this part of $\vk$-space for IR scales.
So, statistically, energy is moved to wavevectors with larger
magnitudes,
%%      but
without much regard for their direction.  Some of
those directions happen to be towards the equal timescale curve, but
plenty of them are also in other directions.
See     Figure~\ref{fig-MHDsimspectrum}.

A special subset of the low $k_z$ fluctuations consists of the
strictly 2D incompressible modes, those having
        $ k_z = 0 $
and     $ v_z = b_z = 0 $.
Recall that purely 2D turbulence, meaning the (triadic) interactions
of a set of strictly 2D modes, forms a closed subsystem with energy
only transferred amongst
the comprising modes.\footnote{In a 3D system,
        the strictly 2D modes do not quite
        form a closed system because a
        nonresonant replenishment process (i.e., driving) can occur;
        see \S\ref{sec:weaktostrong-comm}.}
% \citep{NgBhattacharjee96,
%               VasquezHollweg04-jgr,VasquezEA04,
%               OughtonEA06-2cpt,
%               HowesNielson13,NielsonEA13,DrakeEA13}.
Such
couplings involving only 2D modes
are certainly not attracted to the CB curve.

Some readers may find this discussion of
attraction to the CB curve to be superfluous, arguing that this claim
is not explicit in GS95.
Indeed, it is possible that
this idea originates in a conflation of the symbol ``$\sim$''
with an equality.
On the other hand, given the
variety of interpretations of CB theory that one encounters
in the community, it is perhaps worthwhile to attempt clarification of
this issue.
As an example, we quote    \citet{FormanEA11},
who state           (emphasis ours)
that
\begin{quotation}
    ``When the cascade has the energy dissipation rate
        $\epsilon$
    and also Kolmogorov scaling in the inertial range in the
    perpendicular direction,
        $ v_\perp^3  \sim  \epsilon /k_\perp$,
    CB  \emph{concentrates}
    power at $\vk$ at which
        $ V_A |k_z |  \sim  \epsilon ^{1/3}  k_{\perp} ^{2/3} $.''
\end{quotation}
where they note that $\sim$ means ``goes as'', not equality.
We have attempted to make clear why we disagree with this interpretation
on theoretical grounds.
Perhaps more importantly,
we have seen no evidence, either numerical or observational,
for a concentration of energy on the CB curve (e.g., Figure~\ref{fig-MHDsimspectrum}).

%----------------------------------------------------------

      \subsection{Strong Turbulence as Frustrated
                     Isotropic Spectral Transfer}
           \label{sec:strong-frustrated-iso}

Consider a 3D incompressible MHD system with a mean magnetic
field $\vB_0$.
When $B_0$ is large, the system can be
weakly turbulent with spectral transfer that is
dominantly perpendicular, at least for large enough     $ k_z $.
For moderate (or large)       $ B_0 $,
strong turbulence is usually thought to also involve
strong perpendicular spectral transfer.
However, it may be more helpful to think of strong turbulence as
being
        `frustrated' isotropic transfer,
with the frustration due to weak turbulence modes that are excited by
the strong turbulence.

In greater detail the idea is as follows.
Modes near the equal timescale zone
cascade energy to somewhat smaller scales,
more or less in the isotropic Kolmogorov way
   (since $\tauA$, and $\vB_0$, do not have a dominant
   influence inside the curve).
But some of the modes that are excited/augmented by this (roughly)
isotropic transfer will be in weak turbulence regions of $\vk$-space,
some distance from the zone.
For these,
        %% fluctuations,
the most important nonlinear process involves perpendicular spectral
transfer,\footnote{Mediated
        by the 2D and quasi-2D modes.}
not isotropic transfer
        \citep{ShebalinEA83}.
This moves energy to higher $k_\perp$
and, eventually, into (a different part of) the equal timescale region.
So although some energy `escapes' past the strong
turbulence (equal timescale) boundary, most of it does not do so for very
long.  The perpendicular transfer associated with weak turbulence
fluctuations immediately starts to move the `escaped' energy to higher
        $ k_\perp$,
which is also
towards the equal timescale curve.

One might describe this movement of excitation back towards the
curve as a shepherding or attraction towards the curve.
But, as we have argued earlier in this section,
the weak turbulence perpendicular transfer
is just that---$ \perp$ transfer---and not a pull
towards a region that the transfer process is
unaware of.\footnote{Here we are describing the individual triad
    interactions as ``unaware'' of the equal timescale zone.
    If one considers sets of triad interactions, then in a statistical
    sense there may be nett movement of energy in particular $\vk$-space
    directions.}
The important point is
that the controlling physics is different depending upon whether the
$\vk$-space location is inside or outside the equal timescale curve.

%===============================================================

        \section{The 2D and quasi-2D fluctuations:
                     Correlations, Spectral Power,
                        and Apparent Contradictions}
        \label{sec:corrn-length}

In the previous sections the special roles of the 2D and quasi-2D
fluctuations have repeatedly emerged.
In the context of CB the role of these fluctuations
may be viewed as somewhat formal, for example
in catalyzing strong nonlinear interactions,
eliminating weak turbulence in favor of strong,
and providing a key distinction
between the spectra
of  (steady)  RMHD turbulence and those
envisioned in the GS95 version of CB.
There are also physical phenomena
seen in simulations and observed in the
solar wind, that are most evident in, or most easily explained by,
a healthy admixture of 2D or quasi-2D fluctuations.
 Prominent among these are the frequent
detection of low Alfv\'en ratio\footnote{Clearly,
        the Alfv\'en ratio        $ r_A = E^v / E^b $
        and the residual energy
         $ \sigma_D = (E^v - E^b) / (E^v + E^b) $
        characterize essentially the same property.
        Here $E^{v,b}$
        are the fluctuating kinetic and magnetic energies.}
        (negative residual energy)
inertial range turbulence
        \citep[e.g.,][]{MattGold82-rugged,PerriBalogh10-sigc,
                        BigotEA08-aniso,BigotGaltier11,OughtonEA16},
  %% BigotEA09-aniso, fig 9. BigotGaltier11, fig 3
and the remarkable utility of the 2D phenomenology of magnetic
reconnection in 3D space plasmas
        \citep[e.g.,][]{PhanEA06,RetinoEA07}.
Both of these are best understood in 2D or nearly 2D pictures.
However, there
are frequent arguments given in the CB literature,
usually qualitative in nature, that argue for
the  \emph{absence} of these catalytic quasi-2D
fluctuations. This can be assessed empirically in
numerical models,
as discussed in other sections of this review.
 But there is also a class of
argument that claims to show
on basic physics grounds that 2D fluctuations
are a singular limit, and cannot be present in
real systems. We address those arguments here.

These points were not discussed in GS95, but subsequently it
has sometimes been stated that fluctuations with
%% this timescale ordering
        $ \tauNL $ considerably less than $\tauA$
are not, or even cannot, be present
        \citep[e.g.,][]{MaronGoldreich01,SchekochihinEA09}.
%%.. MG is page 1179, S page 312
Reasoning along these lines has proliferated among practitioners of CB,
often in slightly different versions,
usually citing the previous two
references.
We will address these in some
detail here, and to avoid
ambiguity we begin the discussion with
two relevant quotes:

In \cite{MaronGoldreich01}
        $ k_\parallel $,  $\lambda_\parallel $
are along the  \emph{local}  field,
while $k_z$ is along the global mean field $B_0$.
On their page 1179 it is stated:
\begin{quotation}
  ``Our discussion of intermediate turbulence shows that
        $ \chi = \tauA / \tauNL
               =  v_{\lambda_\perp} \lambda_\parallel / ( V_A \lambda_\perp) $
   increases if it is less than unity. However, it cannot rise above unity,
   since the frequency spread of the wave packets that emerge
   following a strong collision must satisfy the frequency-time
   uncertainty relationship.
   A. Gruzinov (2000 private communication) provides a more physical
   explanation for the upper bound on $\chi $. He points
   out that for $ \chi \gg 1$, two-dimensional motions of scale $\lambda_\perp$
   in planes perpendicular to the local magnetic field are uncoupled over
   separations greater than $\lambda_\parallel /\chi$ along the field
   direction. Thus, during a time interval of the order of
        $ \lambda_\perp / v_{\lambda_\perp}
                \sim
          \lambda_\parallel /v_A  \chi $,
   these motions reduce $\chi $ to order unity.''
\end{quotation}
Offering an interpretation along the same lines,
    \citet[p.~312]{SchekochihinEA09} states
\begin{quotation}
    ``Indeed, intuitively, we cannot have
          $k_\parallel V_A \ll k_\perp u_\perp $:
    the turbulence cannot be any more two dimensional than allowed
    by the critical balance because fluctuations in any two planes
    perpendicular to the mean field can only remain correlated if
    an Alfv\'en wave can propagate between them in less than their
    perpendicular decorrelation time.''
\end{quotation}
Both of these arguments relate to
        \emph{correlations}
and not spectral power, as we shall discuss further below.
The two arguments are also essentially equivalent.
Their essence, restated,
is the physically reasonable observation that
perpendicular planes separated
        (in $\vx$-space)
by large enough $ \Delta z$ in the parallel direction
become uncorrelated in that direction,
because their in-plane
%%        ($k_\perp$)
nonlinear activity is occurring faster than propagation between
planes can convey.
This is, then, fundamentally an argument based on causality.
Correlations cannot be maintained over distances larger
than the range of the fastest signal in a
nonlinear time,
with the fastest signal speed assumed to be the Alfv\'en
speed.\footnote{The
        argument is given in the context of an incompressible model,
        for which the pressure is formally a constraint with
        effectively infinite propagation speed.
        Correlations induced this way are assumed to be negligible.}

The second step of the
argument is that the finite range of parallel
signals
induces development of finer parallel structure,
and thus transfer of energy to $k_z$'s larger than $ 1/ \Delta z$.
The
        %% concomitant
loss of energy at the original low $k_z$ lengthens
the nonlinear time associated with those $\vk$'s,
bringing it closer to their wave timescale, that is,
bringing spectral density closer to the critical balance curve.
Notice that the second
stage of the argument pertains to spectral density and transfer,
not correlations.
%   Hence, in a time of order
%         $ \Delta z / V_A $
%   the system is argued to adjust so that
%         $ \tauNL / \tauA \approx 1 $.

While there may be circumstances where this is the case,
such statements seem hard to justify in general.
Obviously systems can be
initialized with fluctuations satisfying
        $ \tauNL( k) \ll \tauA( k_z) $.
So the more relevant questions are probably to do with the stability,
persistence, and/or dynamical generation of such fluctuations.
Simulation results indicate that as far as the modal energy spectrum
is concerned there is nothing special about the low-$k_z$ values
associated with inertial range fluctuations:
no holes, no jumps, etc, are evident after the system has been able
to evolve for a global nonlinear time or so.\footnote{Excitation of
        the spectrum at $\vk$'s where both $k_z$ \emph{and} $k_\perp$
        are small is typically low, since it occurs via relatively
        weak back transfer processes.  Exceptions occur if conditions
        support significant inverse cascade.
        In either case, the behavior can be masked if the number of
        low-$k_z$ modes available is too small.  For example, in many
        simulations the length of the domain is often (considerably)
        less than ten times the energy-containing scale.}
Indeed, as can be seen in
        Figure~\ref{fig-MHDsimspectrum},
IR spectra are typically smooth as a function of $k_z$ (and $k_\perp$)
in the
        $ \tauNL / \tauA \lesssim 1.5 $
region,
with approximately isotropic contours.
The contour levels for $ \tauNL/\tauA $
indicate that this quantity can certainly be less than unity, even
with the modest Reynolds numbers attainable in simulations.

Notwithstanding the potential realizability of the above two-stage
argument,
one may question the relevance of the quoted
  (\cite{MaronGoldreich01}
   and     \cite{SchekochihinEA09})
statements to the defence of the
        {critical balance theory}  itself.
We will now argue that those two observations, even if true,
do   \emph{not}
support CB theory.
To recognize this requires that one properly
distinguish between the
behavior of     \emph{correlations}
and the behavior of the \emph{spectral density}.

There is a perhaps under-appreciated point
regarding the connection
between spectral power density in non-propagating 2D modes
and the existence of a nonzero parallel
correlation length.
In fact, one cannot have one without the other.

%-----------------------------------------------------------------------------%
\begin{figure} [tbp]
 \begin{center}
    \includegraphics[width=\columnwidth]{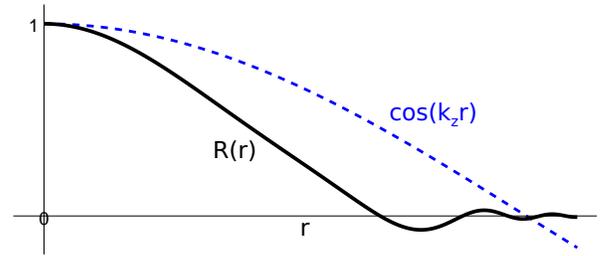}
 \end{center}
  \caption{A normalized correlation function (solid black),
               $ R(r) =
                    \avg{ \vb(\vx) \cdot \vb(\vx + r\zhat) }
                  /  \avg{ \vb \cdot \vb } $,
            and an example Fourier mode $ \cos( k_z r) $
            (dashed blue).
            When the correlation function is essentially zero
            beyond some $r$,
            the integral
                $ \int_0^\infty R(r) \cos(k_z r) \, \d{r} $
            will approach
                $ \ell_c^\parallel = \int_0^\infty R \, \d{r}$,
            as  $ k_z \to 0 $.
          }
  \label{fig-Rb-cosine}
\end{figure}
%-----------------------------------------------------------------------------%
Recall that the parallel spectrum for the magnetic field fluctuations
is essentially the Fourier transform of the correlation function
        $ R_b(r) = \avg{ \vb(\vx) \cdot \vb( \vx + r\zhat) } $,
that is
\begin{equation}
    \Epar_{b}( k_z)
        =
    \frac{1}{2\pi} \int_0^\infty \cos( k_z r) R_b(r) \, \d{r},
  \label{eq:Epar-Rb-FT}
\end{equation}
where the evenness of $ R_b(r) $ has been employed.
The lag $r$ is in the parallel ($\zhat$) direction and the angle
brackets indicate ensemble (or space) averaging.
This relationship makes it clear that, in general, there will be
spectral power at low $k_z$.
Consider
        Figure~\ref{fig-Rb-cosine}
which depicts a normalized correlation function and a cosine
curve of low $k_z$.
One sees that if
        $ R_b(r) \approx 0 $
for large enough $r$, then the product of $\cos(k_z r)$ and $ R_b(r)$
will approach $ R_b(r) $ as $k_z \to 0 $.
That is, at small $k_z$,
    $ \Epar_{b}( k_z) $
becomes approximately constant
        \cite[App.~B]{Lanczos,MattGold82-rugged}.
In particular, for the usual definition of the parallel
correlation length,
%%        $ \ell_c^\parallel $,
%%as the integral of $R_b(r) / R_b(0) $ over all $r$,
        $ \ell_c^\parallel = \int_0^\infty R_b(r) \, \d{r} / R_b(0) $,
one obtains
\begin{equation}
     \avg{ \vb \cdot \vb }  \ell_c^\parallel
  =
     2\pi \Epar_b( 0)
   .
 \label{eq:corr-len-FT}
\end{equation}
Thus when a system has a nonzero parallel correlation length, it
        \emph{must}
have power in the $k_z = 0 $ (i.e., 2D non-propagating)
modes, and vice versa.

One way to construct
a correlation function that is (essentially) zero at large $r$
is to use a superposition of independent Alfv\'en wave packets
   \citep[e.g., GS97,][]{MaronGoldreich01}.
%... above eqn 20 in GS97.  On p 1179 in MG01
If, for example,
the wavepackets have parallel
scale of some specified length
        $ \ell_\parallel $,
and arbitrary $ k_\perp$ structure,
then
        $ R_b(0,0,r) \approx 0 $
for     $ r \gtrsim \ell_\parallel $,
and the associated 1D spectrum is approximately flat for
        $ k_z \lesssim  1 / \ell_\parallel$.

The above discussions clarify the relationship of
the two arguments quoted above to possible defences of
CB theory. It is likely (and indeed almost necessary)
that perpendicular planes become uncorrelated
beyond a certain parallel distance for highly anisotropic turbulence
in the presence of a strong DC mean magnetic field.
However the implication of this finite correlation distance
is not that spectral power is absent in and near the 2D plane,
but rather that such
   \emph{2D spectral power is present whenever the
         parallel correlation scale is nonzero};
    see equation~(\ref{eq:corr-len-FT}).
In our view  a proper interpretation is that the
parallel correlation scale      $ \ell_c^\parallel $
is likely given, in analogy to the above quoted estimate,
as      $ \ell_c^\parallel  \approx  V_A \tauNL $
where the global nonlinear timescale is
        $ \tauNL = \ell_c^\perp / Z $,
for perpendicular correlation scale   $ \ell_c^\perp $
and total turbulence amplitude $Z$.
Meanwhile the reduced 1D parallel magnetic spectrum
flattens to a level
   $ \Epar_{b}( k_z=0)
        =
      \avg{ \vb\cdot \vb} \ell_c^\parallel / 2\pi$
as      $ k_\parallel \to 0$;
see equation~(\ref{eq:corr-len-FT}).
We remind the reader that
        $ \Epar_{b}( k_z=0) $
is the total spectral density in the
2D plane, and quite generally it is nonzero.
This situation is partially consistent with the above quoted passages
from
        \cite{MaronGoldreich01}
and     \cite{SchekochihinEA09}
but here we have clarified
the relationship between correlations and
the behavior of the spectrum between
        $ k_z = 0 $
and     $ |k_z| \sim  1/\ell_c^\parallel $.

We note that our analysis above is actually anticipated by
        GS97
and     \cite{MaronGoldreich01}.
The latter,
referring to the \emph{modal} GS95 spectrum,
our equation~(\ref{eq:GS95spectrum}),
state (p.~1179)
\begin{quotation}
    ``The power spectrum is flat as a function of
        $k_\parallel$
    for
        $ k_\parallel \lesssim k_\perp^{2/3} L^{-1/3} $
    because the velocity and magnetic perturbations
    on transverse scale $k_\perp^{-1} $ arise from independent wave
    packets whose lengths
       $ \lambda_\parallel  \sim  \lambda_\perp^{2/3} L^{1/3}$.''
\end{quotation}
In this quotation $L$ is the outer scale.
As they also note, this requires that
        $ f(u) \approx 1$
for     $ |u| \lesssim 1 $.

The discussion and results presented in this section apply when the
$\vx$-space domain is unbounded
        (so that $\vk$ is a continuous variable)
and the system is statistically homogeneous.
  Appendix~\ref{app:Fseries}
discusses some features of the
  relationship between this case
and the finite domain (or periodic) case for which
   $\vk$
is a discrete variable and Fourier \emph{series} are appropriate.
See also the last paragraph of the Discussion section in GS97.

%===============================================================
%===============================================================

        \section{Non-toroidal Fluctuations}
                \label{sec:otherFlucts}

In GS95, poloidal (e.g., pseudo-Alfv\'en) fluctuations are
discarded at the outset and then later argued to be only weakly
excited at IR scales, by the cascade of toroidal (Alfv\'enic)
fluctuations.\footnote{See the start of
                \S\ref{sec:strong-comm}
        for discussion on why the assumption of strictly
        transverse-polarized Alfv\'enic fluctuations is problematic
        for the
                $\delta b / B_0 \sim 1 $ cases.}
Thus there should be very low levels of parallel variance in such systems.
This may be the case if one restricts attention to IR activity,
because of the strong spectral anisotropy
        ($ k_\perp \gg k_z $)
characteristic of those scales
        \citep[e.g.,][]{PerezBoldyrev08,Beresnyak12-ir,Howes15-not2D}.
However,
for the strong turbulence situation considered in GS95
        (i.e., $ \delta b / B_0 \sim 1 $),
the energy-containing scales are isotropic and
similar arguments predict
that significant excitation of poloidal fluctuations will occur in
less than an eddy turnover time, even when such disturbances are
absent from the initial state.
This is supported by results from simulations of several different
systems using a range of values for $B_0$ and plasma beta ($\pbeta$),
   e.g.,
incompressible MHD, compressible MHD, and hybrid PIC
        \citep{MattEA96-var,DmitrukEA05-rmhd,
                FranciEA15-kinturb,FranciEA15-sw,%
                OughtonEA16,ParasharEA16-aniso}.
These simulations
%% freely decaying simulations
show that when one starts with
purely transverse
initial conditions (at large scales),
a modest level of parallel variance (and hence of poloidal fluctuations)
develops in well under a global nonlinear time.
Levels of $ \sim 5$--20\%
of the total fluctuation energy are reported.
Compressive activity is also seen, typically at somewhat lower levels
of around 10\% for low  $\pbeta$,
indicating that the compressible dynamics is coupled to the toroidal
fluctuations.
Clearly, then, the toroidal (transverse Alfv\'en) modes do not form a closed
system under these types of conditions---which include the GS95 strong
turbulence case of incompressible MHD with
        $ \delta b / B_0  \sim  1 $.

The existence of significant parallel variances at energy-containing
scales means that there is also an additional contribution to parallel
spectral transfer (compared to situations with no parallel variances).
For example,
if parallel variances are present in a system that is otherwise
consistent with the conditions needed for RMHD to be a good
approximation
        (e.g., $\delta b /B_0 \ll 1 $,
         no high-frequency waves in the initial data),
the RMHD model quickly breaks down and
becomes a poor description of the system
        \citep{DmitrukEA05-rmhd}.

What about strictly 2D fluctuations?
These are toroidally polarized since, by definition, they have
        $ k_z = 0 $ and $v_z = b_z = 0 $.
Yet they cannot be propagating Alfv\'en waves because the
associated wave frequency would be
        $ \omega = k_z V_A = 0 $.
Thus from a wave perspective they are anomalous.
In fact, they play a special role from any perspective,
 catalyzing
%%        or mediating
the perpendicular spectral transfer.
See     \S\ref{sec:weaktostrong-comm}.

%===============================================================

%\clearpage
        \section{CB: The View From Observations \& Simulations}
        \label{sec:observations}

In this section we comment on some of the literature associated with
CB in connection with simulations and solar wind observations.  The
papers discussed are selected to indicate the main overall results
obtained to date.
We do not attempt a comprehensive coverage of either the observational
or the numerical results.

\changed{As is the case
  for all IR phenomenologies, critical balance
  approaches assume that a substantial IR exists
   (e.g., at least several decades in wavenumber).
  When the IR is too short it can be very difficult, or even
  impossible, to extract reliable scalings from
  calculated spectra and structure functions.
  For observational analyses the length of the IR is usually not an issue
  since astrophysical and space physics systems often have multi-decade
  IRs.
  However, although $1024^3$ MHD simulations are now routinely achievable
  (and higher resolution runs have been performed
     \citep[e.g.,][]{PerezEA12,Beresnyak14}),
  spectra from many numerical simulations do not exhibit a
  multi-decade IR.  Conclusions on
  spectral scaling laws that are drawn from such studies clearly have
  limitations regarding their applicability to higher Reynolds number
  situations and it behooves us to keep this in mind.
}

        \paragraph{Observations.}
Following early indications
                \citep{SariValley76}
the lengthscale anisotropy of solar wind fluctuations became evident
once two-dimensional correlation functions,
    say $ R( r_\perp, r_\parallel) $,
were constructed from
the observational data
        \citep[e.g.,][]{CrookerEA82,MattEA86b,MattEA90}.
It is well known that the observed
power level in the solar wind magnetic field components
depends on the rotational symmetry of the turbulence,
and on the angle the mean field subtends relative to the radial (flow)
direction,
	$ \theta_{U\!B} $
       \citep{BieberEA96}.
However, evidence that the IR slope of the magnetic power spectrum
depends on
        $ \theta_{U\!B} $
has been presented only more recently
        \citep{HorburyEA08,Podesta09-SWaniso,
                WicksEA10-slopes,WicksEA11,ChenEA11-SWaniso,
                LuoWu10,FormanEA11},
usually based on a \emph{local}
definition of the
mean magnetic field
   (i.e., not the average of the field over the entire interval),
and
employing wavelet methods or structure function methods.
%% average
These studies have typically found
slopes of approximately $ -2 $
for parallel, or nearly parallel spectra
        (e.g., for $ \theta_{U\!B} \lesssim  10^\circ$).
At larger angles $\theta_{U\!B}$  the slope is often close to
        $ -5/3$,
all of which is consistent with a CB interpretation of the spectral
form
        \citep[see, e.g.,][]{HorburyEA12}.

On the other hand, studies that place more stringent demands on the
stationarity of the mean field\footnote{Or its lengthscale
        relative to the lengthscale of the fluctuations.}
used to determine
	$ \theta_{U\!B} $
usually see only a weak variation of slope with
	$ \theta_{U\!B}$
   \citep{TesseinEA09,WangEA16-notCB,TelloniEA19-notCB,WuEA20-notCB},
and have concluded that these observations are not consistent with a CB
model
        (see \S\ref{sec:local-field}).
Indeed,
        \cite{WangEA16-notCB}
state they are not consistent with any known model of MHD turbulence.

An interesting point demonstrated in
        \cite{FormanEA11}
is that the Ulysses observations they consider
are nearly as well fit by the
simple slab+2D model, as they are by a CB one.
Recall that this two-component model
is intended as a kinematic representation of the
anisotropic 3D magnetic autocorrelation
observed in the solar wind
        \citep{MattEA90},
and not as a dynamical model.
Naturally, no one really believes that
        $ E^v(\vk) $
and     $ E^b(\vk) $
are zero everywhere except along the
        $ k_z $ axis
and in the
        $ k_z = 0 $ (2D) plane;
see Figure~\ref{fig-MHDsimspectrum}.
The point is that this skeletal `stick figure' description provides a
reasonably accurate model that is simple to work with,
in particular for scattering theories
        \citep{Shalchibook}.
A ``quasi-2D'' and ``quasi-slab'' model,
in analogy to the models of     \cite{Schlickeiser},
would be more reasonable,
widening the respective $\delta$-functions into narrow regions
around the respective axes of symmetry. In this regard it
is intriguing to note that the CB model, roughly speaking,
resembles such a quasi-2D model (possibly with removal
of the power along the strictly 2D plane).
However, because it imposes
        $ f(u) \ll 1 $
for     $ u \approx 0 $,
the CB model fails to include sufficient power in the
parallel or quasi-parallel wavevector modes (say, quasi-slab)
to efficiently scatter cosmic rays and SEP
protons with energies between 10\,keV and a few GeV
        \citep{Chandran00}.
These are particles that have cyclotron resonant wavenumbers
corresponding to the inertial range of solar wind turbulence.
Their mean free paths provide an additional constraint on theoretical
descriptions of the turbulence spectrum, seemingly requiring
such parallel resonances
        \citep[see, e.g.,][]{BieberEA94}.
Explaining this apparent resonant scattering remains a serious
challenge for CB theory (see e.g., \cite{LynnEA13}).

Another interesting prediction of CB theory relates to
a change of slope in the parallel wavenumber spectrum at
higher $k_z$.
In particular when the perpendicular
wavenumber exceeds the dissipation (breakpoint) wavenumber
        $ \kdiss = 1 / \ldiss$,
those scales are beyond
        %% the high-$ k$  end of
the inertial range and in a region where the spectrum steepens.
According to CB there is a corresponding parallel wavenumber
        $ k_z L  = (\kdiss L)^{2/3}
                 = (L / \ldiss)^{2/3}  $,
beyond which
one expects a steepening in the parallel wavenumber reduced spectrum,
        $ \Epar(k_z) $.
For typical\footnote{For example,
        using nominal 1\,AU values of $ L \approx 10^6\, $km
        and  $ \ldiss \approx 100\, $km
        yields $1/k_z \approx L/500 \approx 2000\, $km.}
solar wind parameters this steepening
should be apparent at a mid-inertial range scale
  (i.e., considerably larger than the scale at which the perpendicular
         spectrum steepens, $ \ldiss $)
if the CB spectral form is correct.
No such steepening has been reported as far as we are aware
        \citep{DuanEA18}.
Although the solar wind is an extremely useful ``natural laboratory,''
one should nonetheless keep in mind that
it is not a perfect system for verifying, or
not, predictions of CB-based (MHD) models
since there are complications due to expansion
effects and possible non-axisymmetry features
      \citep[e.g.,][]{VerdiniGrappin15,VerdiniGrappin16,VerdiniEA19-3Daniso}.

   \paragraph{Simulations.}
Numerous simulation studies have examined spectra and/or structure
functions and have reported
at least tentative support for the
scalings associated with GS95 type spectra or the Boldyrev
        (\citeyear{Boldyrev05,Boldyrev06})
  %% anisotropic
modification of CB.
Here we highlight a few of these, making no attempt to be
comprehensive.
We do note that there have been some contentious claims and that one
should keep in mind that even the highest resolution simulations
performed only have modest Reynolds numbers compared to most
space physics and astrophysics systems.

   \cite{ChoVishniac00-aniso}
presented what was perhaps
the first simulation support for CB scalings.
For incompressible 3D MHD\footnote{Note that these,
        and many other 3D incompressible full MHD simulations,
        retain all poloidal fluctuations.
        However, in GS95 these are discarded so that,
        strictly, the comparison with simulation results should only
        be to quantities derived from the toroidal fluctuations.
        The RMHD approximation eliminates poloidal fluctuations, but
        is based on a strong (not order unity) $B_0$.}
with
        $ \delta b / B_0 \approx 1 $
they found that, mostly,
        $ \ell_\parallel  \propto  \ell_\perp^{2/3} $,
where these lengthscales were determined with respect to a
(two-point) estimate for the
        \emph{local} mean
magnetic field
  (i.e.,  not relative to the global field).
This is apparently also the
first paper to emphasize the importance of using a local mean field to
obtain CB scalings, as discussed in~\S\ref{sec:local-field}.
In some cases the scaling was closer to $1/2$ than $2/3$,
which may be consistent with Boldyrev's
        (\citeyear{Boldyrev05,Boldyrev06})
extension of CB.

The issue of ``CB scaling''
i.e., $ \ell_\parallel  \propto  \ell_\perp^{2/3} $,
warrants further scrutiny due to a certain
level of imprecision both in its definition and in
its interpretation.
It was already pointed out in   \S\ref{sec:rmhd}
that the scaling with $k_\perp$
of the parallel bandwidth of the
spectrum in RMHD is
essentially the same as what is sometimes
called CB scaling. This ``RMHD scaling''
was demonstrated in     Eq.~(\ref{eq:rmhd-scaling}),
in a simple calculation carried out with respect to a fixed
mean field.
Related calculations employing
structure functions computed relative to a local mean field
        \citep{ChoVishniac00-aniso,MalletEA15}
generally find greater anisotropy
than that found using a global mean field
        \citep{MilanoEA01,MattEA12-local}.
This effect
has been attributed to higher-order statistics and intermittency
        \citep{MattEA15-philtran}.
It is clear that
analyses employing
global and local mean fields
may sometimes differ at a significant
level. However it is equally clear that
such differences
must be  \emph{small}
when $\delta b/B_0$ is small.
Therefore in an RMHD regime,
where the the mean field must be strong,
results based on
local mean fields and global mean fields
must very nearly coincide.

There is another interesting technical difference
that sets apart certain
local structure function studies
        \citep{ChoVishniac00-aniso,MalletEA15}
from standard fixed mean field direction
analyses that lead to        Eq.~(\ref{eq:rmhd-scaling}).
In particular   \citet{ChoVishniac00-aniso}
introduce the
idea of following a contour of constant second-order structure
function
        $ S(\ell_\perp,\ell_\parallel) = S_0 = const. $
until it encounters the
        $ \ell_\parallel = 0$ axis
at      $ S( \ell_\perp^0,0) = S_0 $
and the $\ell_\perp=0$ axis at $S(0,\ell_\parallel^0)=S_0$.
Significance is then attributed to the
pair $(\ell_\perp^0,\ell_\parallel^0)$.
With allowance for our notation,
the authors state [above their Eq.~(18)]
that
\begin{quotation}
  ``The ($\ell_\perp$)-intercept and ($\ell_\parallel$)-intercept
    of a given contour can be regarded as a measure of
    (average $k_\perp$) and (average $k_\parallel$)
    for the corresponding eddy scale.''
\end{quotation}
To the extent that one may make the approximate identification
of $\ell_\perp$ with $1/k_\perp$
and $\ell_\parallel$ with $1/k_\parallel$,
this statement is certainly true.
However one may readily show that this ratio measures the relative
strength of two different reduced energy spectra,
one parallel and one perpendicular.
It is quite plausible that this provides a measure of the aspect
ratio of some eddies, which was its original intent.
Nonetheless it is difficult to see how
one may relate this ratio to the spectral
characteristic inherent in the critical balance wavenumber
condition, embodied in
        Eqs.~(\ref{eq:CB-condit})  and  (\ref{eq:Higdon-k}).

%Cho \& Vishniac said [p 280]:
% ``The R-intercept and z-intercept of a given
%   contour can be regarded as a measure of k_\perp and k_\parallel
%   for the corresponding eddy scale.''

  \cite{ChoEA02}
performed forced 3D incompressible MHD spectral method
simulations with moderate $B_0$.
Calculating the spectral transfer (aka cascade) timescale,
        $ \tauS $,
using only local contributions to the
nonlinear terms,\footnote{Band-limited
        in Fourier space.}
they obtained the IR scaling
        $ \tauS  \approx  k_\perp^{-2/3} $,
which is consistent with a CB spectrum.
           \remarkHide{bottom right of their p564}
As noted in
        \S\ref{sec:strong-comm}
they also used these results to assess the suitability of fitting
Gaussians, exponentials, step functions, and Castaing functions
to the shape function
       $ f ( u ) $
in
        Eq.~(\ref{eq:GS95spectrum}),
concluding that an exponential form was a good choice.

Ostensibly
        \cite{MaronGoldreich01}
presents some simulation support for the GS95 strong
turbulence spectral form, and also for
        $ \tauNL / \tauA   \approx  1 $
holding in the IR.
However, although they perform 3D incompressible MHD simulations
they impose
        $ \delta b /B_0 \approx 1/300 $
which does not conform to the
assumptions of the GS95 strong turbulence model
        (for which  $ \delta b /B_0 \approx 1 $).
Thus any
consistency with GS95 might require a distinct explanation.
In fact, the smallness of $\delta b$ means that their simulations are
very close to being RMHD ones,
except that the poloidal
        (pseudo-Alfv\'en)
fluctuations
are small rather than excluded.
Considering the results, they find that the
perpendicular inertial range spectrum,
        $ E(k_\perp) \sim k_\perp^{-\alpha} $,
has
        $ \alpha \approx 3/2$
rather than the GS95 value of   $ 5/3$.
It was suggested this might be due to intermittency effects,
but it is also consistent with
        Boldyrev's
        (\citeyear{Boldyrev05,Boldyrev06})
modification of CB.
Recall that
        \cite{ChoVishniac00-aniso}
also saw this scaling in some, but not all, of their simulations.

The question of whether the perpendicular energy spectrum scales
as
        $ \sim k_\perp^{-5/3}$,
as in GS95,
or as
        $ k_\perp^{-3/2} $
which could occur if there is scale-dependent alignment of the
        $ \vv $
and     $ \vb $
fluctuations
        \citep{Boldyrev05,Boldyrev06},
has also attracted interest
        \citep[e.g.,][]{MasonEA06,MasonEA08,MasonEA12,
                PerezEA12,PerezEA14,Beresnyak14}.
Much larger simulations may be required to settle this
point.
It is also worth noting that the Kolmogorov-style expectation of
powerlaw scaling in an inertial range applies to quantities that are
conserved
  (in the absence of dissipative effects).
So there is a  basis  for
expecting to see a powerlaw IR for the total energy spectrum
(or the Elsasser energy spectra),
but not for the kinetic or magnetic energy spectra,
since neither kinetic energy nor magnetic energy is
separately conserved.\footnote{See, however,
                \cite{BianAluie19}
        for discussion of when the cascades of kinetic and magnetic
        energy can become decoupled.
%%        See also \cite{YangEA17-compcascade}.
        }

There is another possible explanation for a     $ k_\perp^{-3/2} $
IR spectrum in simulations with a strong        $ B_0 $,
one that apparently has not been explored in this context
        \citep[cf.][]{ZhouEA04,ZhouMatt05}.
If the 3D simulation is actually close to RMHD conditions
        \citep[e.g.,][]{MaronGoldreich01}
there is a possibility of significant build up of magnetic energy
in the large lengthscale 2D modes ($k_z = 0$ and $k_\perp$ small),
with amplitude $B_1$, say
        \citep{DmitrukMatt07}.
If      $ k_\perp B_1 $ is large enough,
then    \emph{in-plane}
propagation at the
        $ B_1 $-based  Alfv\'en speed
may control the lifetime of triple correlations
through associated propagation effects,
instead of the in-plane nonlinear time.
 %%   (the latter, by CB, approximately equal to the parallel,
 %%      $B_0$-based, Alfv\'en time.)
In such cases Kraichnan's
        (\citeyear{Kraichnan65})
original reasoning may apply:
for perpendicular transfer one argues that the cascade rate is
        $ \epsilon = \tau_3(k_\perp) E^2(k_\perp) k_\perp^4 $,
so that taking  $ \tau_3 = 1 / (k_\perp B_1) $
immediately leads to
        $ E(k_\perp) \sim k_\perp^{3/2} $.
Note that in this situation,
CB need not be invoked to obtain this spectral law.

Most of the discussion in this section has related to GS95's second
scenario, in which the large-scale turbulence starts in a CB state.
Simulation support for their first scenario,
wherein an initial weak turbulence system evolves and develops
smaller perpendicular scales that are strongly turbulent,
has also been presented.
The weak-to-strong transition has been reported for
shell-model RMHD
        \citep{VerdiniGrappin12}
and for full 3D incompressible MHD
        \citep{MeyrandEA16}.

%=============================================================

        \section{Summary and Conclusions}
                \label{sec:summary}

In classical hydrodynamic theory for
an isotropic incompressible stationary turbulence, there is only one
relevant dimensionless number and one relevant timescale
at high Reynolds number.
For magnetohydrodynamics or plasmas,
there are more available timescales and controlling
parameters, and possibly even many    \citep{WanEA12-vKH}.
Models such as Critical Balance and Reduced
MHD represent attempts to achieve simpler descriptions
by collapsing some of these complications,
exploiting anisotropy,
reducing dimensionality, eliminating timescales, and so on.
The essence of CB is the reduction of
relevant independent timescales by equating the
values (and some of the roles) of the nonlinear timescale and the
Alfv\'en timescale. Here we have attempted to
evaluate the efficacy and generality of the
CB approach by examination of its derivations, its implications, and its
relationship to other models.
Of special interest is the relationship
between CB
and RMHD        \citep{OughtonEA17-rmhd}
given that their origins are quite different---CB being essentially
a dynamically emerging state, while RMHD
is a dynamical model of evolution.
Nevertheless these two models ultimately are applied
to similar circumstances of highly anisotropic MHD turbulence
in a regime in which compressibility is unimportant.

Even though CB and RMHD are implemented as
descriptions of turbulence, each possesses interesting relationships to
wave modes that exist within their purview.
Of course, turbulence is much more than a collection of interacting
waves and indeed may not have much in common
with linear modes at all.\footnote{We note
        that the notion that
        kinetic Alfv\'en wave turbulence,
        or whistler turbulence,
        or indeed any wave-mode turbulence are in fact turbulence can be
        questioned.
        The point is that for these wave (aka weak) ``turbulence''
        systems, the nonlinear dynamics is far
        from being dominant because the nonlinear timescale is much
        longer than the wave timescale.
        Thus, it is unclear why these systems would be analogous to
        strong turbulence systems for which the wave timescale is weak
        or at best comparable to $\tauNL$.
}
For example, at any instant it is mathematically valid to
decompose a turbulent state into a sum of linear eigenmodes by
projecting the spatial structure
onto such a basis.
However, this implies neither
that the physics is of this nature
     (cf.\ projecting a spherical wave onto a plane wave basis),
nor that the expansion coefficients are even approximately stable
in time
     (e.g., the coefficients obtained from decompositions performed at
           slightly different times could be quite different).

In some ways
CB lives close to the world of waves given its premise that
measures of nonlinearity and measures of wave activity
emerge as being of equal numerical strength.
In other ways
it acknowledges a secondary relevance of the wave physics with
statements like
\emph{``the assumption of strong nonlinearity implies that wave
  packets lose their identity after they travel one wavelength
  along the field lines.''}
        \citep{ChoVishniac00-aniso}.  \remarkHide{(p 275)}
A clean interpretation of the physics of CB has been provided by
        \citet{Terry18},
who argues that it is best understood as a hypothesis regarding
the fastest timescale influencing the lifetime of the (nonlinear)
triple correlations.

%%     \emph{``interactions are so strong that a  `wave packet'
%%         lasts for at most a  few wave periods''}  (GS95).
%% (their footnote 2).

It is also the case that
turbulence and waves are not antithetical or mutually exclusive.
They may, however, be associated with different frequency and/or
lengthscale ranges
        \citep[e.g.,][]{DmitrukEA04-discrete,DmitrukMatt09,
                        AndresEA17,CerretaniDmitruk19}.
In particular, the turbulence can be of relatively low frequency
  (for the energy-containing scales)
and the wave activity at relatively high frequencies.
For example, suppose that you have hiked to the top of a hill
and the wind is very gusty there
(or, in turbulence language, the velocity fluctuations are
 strong and irregular at the energy-containing scales).
It is still relatively easy to have a conversation in such conditions,
at least over distances of a few metres,
because small-amplitude high-frequency sound waves
can propagate well enough.

Some have argued that when fluctuation amplitudes
have  polarization  similar to
small amplitude wave modes, the turbulence is
        \emph{de facto}
wave-like.
However, this polarization condition seems to us too weak a
constraint for the conclusion.  The wave timescale can be
relevant without the fluctuations being waves.  Indeed, this
is essentially how GS95 present CB.  They note the waves last at most a
few periods (because nonlinear effects are strong), meaning that such
fluctuations are far from being waves in the usual sense of coherent
disturbances that propagate many wavelengths
and have recognizable dispersion relations.
This issue remains controversial.
For example,
        \citet[][their \S4.5,4.6]{NazarenkoSchekochihin11}
make relevant comments.
Assuming that an initial wave packet of linear waves is present,
they ask
    how much the linear wave polarization of fluctuations
    can be preserved in a strongly turbulent nonlinear state?
The answer, they suggest, is that there is a
      tendency (of the cascade dynamics) to preserve the (linear) wave
      polarization to the maximum possible degree.
This is referred to as the
        \emph{polarization alignment conjecture}
and they note it is analogous to an argument presented in
        Boldyrev's   (\citeyear{Boldyrev06})
extension of CB.

Similarly
        \cite{HowesNielson13}
show that when waves are present this may be difficult to verify via
dispersion relations
because the changing background in which they propagate
makes the dispersion relations rather featureless.
They suggest that instead the eigenfunction structure
        (i.e., the polarizations)
could be used to identify the presence and/or importance of waves.
These discussions appear to be somewhat at odds with the
original statement of CB, wherein
GS95 state that `waves' last at most a period or so.

Discussions of applicability as well as correct application of
CB have often revolved around the definition
of ``mean field'' and in particular the issue of
whether the
Alfv\'en propagation effect should always be discussed in terms
of a \emph{local} mean field.
This issue becomes of central importance
when arguing that the mean field
is \emph{always} strong at small scales in the inertial range,
and therefore the turbulence can essentially
always be viewed locally as obeying RMHD.
In this review we have not delved into this claim in any detail,
and we remain unsure that conditions for deriving RMHD
are realizable in the random local and dynamic
coordinate systems that would be required
        \citep[see][]{OughtonEA17-rmhd}.
We note, however, that if the classical view of RMHD is adopted, then
the global mean magnetic field direction and the total magnetic field
direction cannot be very different
        (as   $ \delta B/B_0 \ll 1 $   is required)
and therefore the argument concerning the role of local fields
is relegated to a matter of lesser consequence.

After examining the
history and applications of CB,
we have arrived at the opinion that the relationship between
RMHD and
        the GS95 version of  CB
may be compactly summarized by a few similarities
and a few important differences.
Recognizing the similarity of these two approaches
may permit the advantages of each to be more readily exploited rather
than emphasizing distinctions that are difficult to
explore.
Some of the main features of CB---incompressibility,
        the polarization of the fluctuations,
        the anisotropic spectrum,
        and the Higdon curve---are
also properties of RMHD.
But there are other features of CB (as formulated in GS95)
that are not present in RMHD:
 an attraction to the Higdon curve,
 the lack of very low frequency quasi-2D fluctuations,
 the restriction to   $ \tauNL   \approx  \tauA $
    (versus RMHD's    $ \tauNL \lesssim  \tauA $),
 and the possibility of a derivation without assuming a strong mean
 field.
Among these properties that distinguish CB and RMHD,
all except one are more restrictive in CB, so that
one might judge RMHD to be the less restrictive,
or ``bigger,'' theory.
On the other hand, RMHD requires a strong mean field, whereas
        (the GS95 version of) CB
explicitly imposes a mean field of only moderate strength.
This suggests that it is RMHD that is a more restrictive theory with,
in some sense,
a ``smaller'' domain of applicability.
Thus, paradoxically, RMHD seems to be both a more restrictive theory
than CB and a less restrictive one.
To arrive at a clear conclusion regarding this issue requires
an immersion in the conceptual bases of the two models and their
respective derivations,
which we have attempted here for CB, and for RMHD
in    \citet{OughtonEA17-rmhd}.
A complementary approach is to examine carefully
numerical performance of each model in the context of full MHD solutions.
This too has been touched on here, and is dealt with in more detail
elsewhere
        \citep{DmitrukEA05-rmhd, GhoshParashar15-i, ChhiberEA20-tauNL}.
We can only trust that sufficient evidence will be developed and
recognized so that the community will eventually develop an
  accurate
consensus on  these issues.
Our intention in this paper has been to contribute to that factual basis.

\changed
{Looking to the future,
  it is apparent that there are still issues connected with CB that
  are yet to be fully resolved.  Some of these have been discussed
  above. By way of a summary we list the main ones here:
   \begin{itemize} \itemsep=0.5ex
     \item
       Why should the small-amplitude Alfv\'en wave mode (with its
       toroidal polarization) be a suitable basis for developing a
       theory of \emph{large}-amplitude fluctuations that need not be
       toroidal?
       As is well-known, solar wind observations suggest magnetic
       fluctuations are more typically associated with polarizations
       such that the total field lies on the surface of a sphere:
       $ |\vB_0 + \vb| = $ constant.
     \item
       What are appropriate conditions to place on the (spatial
       and/or temporal) stability of local mean fields?  Different
       scaling results seem to emerge when different conditions are
       imposed.
     \item
       Solar wind observations do not indicate
       steepening of the parallel spectrum at a smaller $k$ than for
       the perpendicular spectrum.  This is in conflict with the
       spectral anisotropy of a CB model.
     \item
       In strong MHD turbulence the Alfv\'en timescale is certainly
       relevant.
       To what extent are properties of (linear) Alfv\'en waves also
       relevant (e.g., polarizations)?  Concrete suggestions have been
       made, such as the polarization alignment conjecture,
       but a full understanding is still being sought.
   \end{itemize}
}

\changed{Finally,
   we note that two recently launched spacecraft,
    \emph{Parker Solar Probe} and \emph{Solar Orbiter}, are exploring
   regions that are closer to the sun than any other spacecraft have
   ventured
     (see \cite{Neugebauer20} and the other papers in that same
     special issue).
   It will be fascinating to see what
   observational results from these missions reveal
   about the applicability of CB in such regions of the solar wind,
   particularly as a common assumption is that Alfv\'en wave activity
   will be more pronounced there.
   We await these studies with interest.
}

%------------------------------------------------------------------------------%
\begin{acknowledgments}
  This research was supported in part by NASA Heliospheric Supporting
  Research grants NNX17AB79G, 80NSSC18K1210, and 80NSSC18K1648,
  and the Parker Solar Probe mission though the ISOIS project and
  subcontract SUB0000165 from Princeton University.
\end{acknowledgments}

%------------------------------------------------------------------------------%
%% \clearpage

\appendix

        \section{A: Different Notions of Alfv\'enic Turbulence}
          \label{app:Alfvenic}

  \emph{Alfv\'enic turbulence}
is used in several different ways in the literature.
In the present context, Alfv\'enic might
        be defined loosely
to mean that the turbulence exhibits
        features---perhaps only vestigially---that are not inconsistent
        with those of Alfv\'en waves.
However this term has also been used to refer to
diverse circumstances that are only
incidentally related to one another.
Here is a list, probably not exhaustive, of some of these:

  \noindent $\bullet$ \emph{High cross helicity turbulence.}
When the cross helicity is large, e.g.,
 $ |\sigma_c| \to 1$
(where
 $ \sigma_c = 2 H_c/E $)
incompressible MHD approaches a state that resembles a large amplitude
Alfv\'en wave
        \citep{Parker-cmf}.

  \noindent $\bullet$ \emph{The incompressible small-amplitude MHD wave}
is the Alfv\'en mode.
This is, possibly, a reason that
\emph{incompressible MHD turbulence} is sometimes called
``Alfv\'enic'' regardless of its cross helicity

  \noindent $\bullet$ \emph{Large amplitude Alfv\'en wave.}
In these large amplitude propagating states the nonlinearity is cancelled out.
Defined originally in incompressible MHD, these modes may also survive
for moderately long timescales in compressible cases
  \citep{Barnes79a,PezziEA17-headon,PezziEA17-jpp,PezziEA17-apj}.

  \noindent $\bullet$ \emph{Equipartitioned kinetic and magnetic energies.}
If the energy densities for the fluid-scale velocity fluctuations and
the magnetic fluctuations are approximately equal (unit Alfv\'en
ratio), then the state of the turbulence resembles Alfv\'en waves in
that regard. This is sometimes called an ``Alfv\'enic state,''
and might imply either a global condition
        $ (\delta v)^2  \approx  (\delta b)^2$,
or one that holds over a range of scales,
        $ |\delta v(k)|^2  \approx |\delta b(k)|^2 $.

  \noindent $\bullet$ \emph{Transverse turbulence}.
The polarizations
of the magnetic and velocity field fluctuations are called
        \emph{transverse}
when they have no projection onto the local mean magnetic field
direction.
This property is shared with small amplitude Alfv\'en waves
 (more precisely, the latter are \emph{toroidally} polarized).

%\noindent $\bullet$ {Dynamic Pressure balanced structures}.

%------------------------------------------------------------------------------%
       \section{B: Perpendicular Spectral Transfer via 3-mode Resonance}
        \label{app:Shebalin}

The weak turbulence explanation for strong perpendicular spectral
transfer was first presented in
        \cite{ShebalinEA83}.
See also
        \cite{Bondeson85}, \cite{Grappin86}, and \cite{OughtonEA94}.
Here we briefly outline the steps in the argument.

The system considered is 3D incompressible MHD with a uniform mean field
        $ \vB_0 = B_0 \zhat $,
and an initial condition that is a superposition of linearized
solutions to the equations.
 %% In general this would include Alfv\'en waves, pseudo-Alf\'en waves,
 %% and 2D structures (these have $k_z = 0$).
Perturbation theory is employed to calculate nonlinear corrections to
the leading-order solutions.
Using Elsasser variables the solutions can be written as
        $\vz^\pm(\vk, t) = \va^\pm_{\vk} (t)
                          \exp\left(
                                i [\vk \cdot \vx - \omega(\vk)t]
                              \right) $,
with    $ \va^\pm_{\vk} $
slowly-varying amplitude functions.
The Alfv\'en dispersion relation is
        $ \omega( \vk) = \pm \vk \cdot \vB_0 $,
with the sign convention that the plus sign is associated with
        $ \vz^- $ modes,
and the minus sign with
        $ \vz^+ $ modes.
Hence, the modes propagating parallel (anti-parallel)
to $\vB_0$ are of the
        $ \vz^- $   ($ \vz^+ $)
type.
In general, non-propagating modes with $k_z = 0 $,
are also present
        \citep[e.g.,][]{ShebalinEA83,MontMatt95}.

Because the nonlinear term is quadratic, the simplest possibility for
resonant interaction is that two distinct linearized solutions couple
to drive a third linearized solution,
often called a ``3-wave resonance'' although the linearized modes are
not necessarily waves.
As an example consider the interaction of a positive cross helicity
mode
        $ \vz^+( \vk_2, t) $
with a negative cross helicity mode
        $ \vz^- ( \vk_1, t) $
to drive another negative cross helicity mode
        $ \vz^-( \vk_3, t) $.
The governing equation is
        $ \partial_t \vz^-(\vk_3, t)
                \sim
          - \vz^+(\vk_2,t) \cdot \nabla \vz^-(\vk_1, t) $,
where only the nonlinear term is shown.
The usual frequency and wavevector matching conditions arise,
corresponding to conservation of energy and momentum,
\begin{eqnarray}
   \omega(\vk_1) + \omega(\vk_2) & = & \omega(\vk_3)
   \quad \Rightarrow \quad
      k_{1z} - k_{2z} = k_{3z},
  \qquad\quad
 \label{eq:om-matching}
\\
   \vk_1 + \vk_2 & = & \vk_3
   \qquad\; \Rightarrow \quad
      k_{1z} + k_{2z} = k_{3z}.
 \label{eq:k-matching}
\end{eqnarray}
Subtraction shows that $ k_{2z} = 0 $
so that the advecting mode        $ \vz^+(\vk_2,t) $
is actually a 2D mode and not a propagating linear wave
at all.\footnote{Hence,
        for incompressible MHD `3-\emph{wave} resonance' is
        a misnomer, albeit a well-used one.}
It is nonetheless a valid solution of the linearized MHD equations
        \citep{MontMatt95},
with the physical interpretation of a coherent structure that varies
across $\vB_0$ but not along it.  As far as nonlinear effects are
concerned, the relevance is that a propagating mode interacts with the
2D mode for many wave timescales,
        $ \tauA \approx 1 / |\omega(\vk_1)| $,
and, specifically, does so for at least a nonlinear time.
(This avoids the chopping of nonlinear effects produced during
non-resonant wave-wave interactions.)
Moreover, because $k_{2z} = 0 $,
the transfer of energy from
        $ \vz^-(\vk_1) \to \vz^-(\vk_3) $
occurs at fixed $ k_z $ and
is thus strictly perpendicular transfer, at this order.
From
        Eq.~(\ref{eq:k-matching})
the $k_\perp$ transfer satisfies
        $ | \vk_{3\perp}| = |\vk_{1\perp} + \vk_{2\perp}| $,
and will typically be towards higher $k_\perp$.
%------------------------------------------------------------------------------%

        \section{C: Simulation Details}
        \label{app:sims}

The data used to produce Figure~\ref{fig-MHDsimspectrum}
was obtained from a 3D incompressible MHD Fourier pseudospectral
method simulation,
of resolution $ N^3 = 1024^3$
in a periodic box of dimensionless size
        $ 2 \pi $.
The (Cartesian) components of the wavevectors are thus integers in the
range $ -511 $ to $512$,
and
the  longest allowed wavelength in the periodic box
corresponds to a wavenumber of unity.

A parallel (MPI-based) algorithm is employed to solve standard
dimensionless equations,
\begin{eqnarray}
  \PD{\vv}{t}  & = & - \nabla p^*
                     + \vv \times \vct{\omega}
                     + \vj \times \vB
                     + {\nu} \nabla^2 \vv,
\\
  \PD{\va}{t}  & = &   \vv \times \vB
                     - \nabla \Phi
                     + \eta \nabla^2 \va   .
   \label{eq:mhd}
\end{eqnarray}
Here
        $ \vv( \vx, t) $ is fluctuating velocity,
        $ \vct{\omega} = \nabla \times \vv $ the vorticity,
        $ \va( \vx, t) $ the vector potential for the magnetic
        fluctuations $\vb = \nabla \times \va $,
        $ \vB = \vB_0 + \vb $  is the total magnetic field,
        $ \vj = \nabla \times \vb $  is the electric current density,
        $ p^* $ is the total (fluid plus magnetic) pressure,
and     $ \Phi $ a gauge function used to enforce the Coulomb gauge on
        $\va$
     \citep{CanutoEA,GhoshEA93-cpc}.
Time advancement is via a second-order Runge--Kutta method

The initial conditions are chosen to be consistent with the GS95
assumptions.
This includes a strong turbulence energy partitioning
        $ \delta v = \delta b = B_0 = 1$,
and
purely toroidal fluctuations
  (i.e., excited modes are polarized parallel to $\vk \times \vB_0$)
with wavevector magnitudes in the range
        $ 3 \le | \vk | \le 7 $.

The amplitude of each excited wavevector mode follows the spectral
shape
        $  1/ \sqrt{ 1 + (k/K_0)^q}       $,
where the ``knee'' $ K_0 = 3$
and $q = 2 + 5/3$
means the omnidirectional spectrum has a Kolmogorov $k^{-5/3}$ powerlaw at
high $k$.
Phases of each Fourier mode are set using Gaussian random variables so
that correlations amongst modes are small.
In particular, the net cross helicity and magnetic helicity were both
close to zero.

The timestep was $ 2.5 \times 10^{-4} $ and
viscosity and resistivity were set to
        $ 7.3 \times 10^{-4} $
so that the initial (kinetic and magnetic) Reynolds numbers are
        $ \approx 500 $.
This combination ensures that the cutoff wavenumber
        ($ k_\text{wall} = N/2 = 512 $)
is at least triple the maximum Kolmogorov dissipation wavenumber,
        $ \kdiss( t) $,
a criterion that is important for obtaining accurate higher-order statistics
    \citep[e.g.,][]{DonzisEA08,WanEA10-accuracy}.
The spectrum displayed in
        Figure~\ref{fig-MHDsimspectrum}
was calculated just after the time of
maximum dissipation rate when the large-scale Reynolds numbers were
        $ \approx 300 $.

%==================================================================

        \section{D: CB in Finite Domains: Fourier Series}
        \label{app:Fseries}

The arguments presented in
        \S\ref{sec:corrn-length}
apply when Fourier space is
a continuum, as occurs for unbounded $\vx$-space domains.
In numerical work the spatial domain is typically bounded
and thus     $\vk$-space
is discrete,
e.g., for spectral method simulations.
In that case
        Fourier \emph{series}
provide an appropriate representation.
This leads to a different $\vk$-space form for the correlation length
that involves intricacies arising from periodicity effects
        \citep{Lanczos,MattGold82-rugged}.

Let us write the
magnetic fluctuation as a truncated Fourier series, where $x,y,z$ are
each over the interval $ [0, 2\pi ) $, and the wavenumber components
range over the integers from $-N$ to $N$:
   $ \vb(\vx)
                =
      \sum_{\vk}  \tilde{\vb}_{\vk} \mathrm{e}^{i \vk \cdot \vx}
   $.
The correlation function is
\begin{eqnarray}
   R_b(r)
       & = &
     \avg{\vb(\vx) \cdot \vb(\vx + r\zhat) }  %%_\text{ensemble}
 \nonumber
\\
       & = &
    \sum_{k_1 = -N}^{N} \sum_{k_2 = -N}^{N} \sum_{k_3 = -N}^{N}
      \mathrm{e}^{i \vk \cdot \vr} S(\vk)
   ,
  \label{eq:R-FS}
\end{eqnarray}
where
        $ \sum_{\vk} $
is shorthand for the triple sum shown in
        Equation~(\ref{eq:R-FS}),
   $
    \avg{ \tilde{\vb}_{\vk} \cdot \tilde{\vb}^{*}_{\vk'}}  %%_\text{ensemble}
        =
    S(\vk) \delta_{\vk, \vk'}
   $,
with $*$ denoting complex conjugation,
and
   $ S(\vk) $
is (twice) the energy spectrum.
Integrating the correlation function up to some temporarily
unspecified limit $X$ yields an expression for the parallel
correlation length in this discrete $\vk$-space case,
\begin{eqnarray}
   \avg{\vb \cdot \vb} \ell_{3}
        & = &
     \int_{0}^{X} R_b(r) \, \d r
 \nonumber
 \\
        & = &
      2X \Ered( 0)
        +
      4 \sum_{k_3 = 1}^{N} \Ered( k_3) \frac{\sin(k_3 X)}{k_3} .
  \label{eq:corr-len-FS}
\end{eqnarray}
Here
        $ \Ered( k_3)
                =
          \frac{1}{2}
          \sum_{k_1,k_2}  S(k_1, k_2, k_3)
        $
is the reduced energy spectrum.
The choice $ X = \pi $ (half the domain width)
makes the formula for
        $ \ell_3 $
formally equivalent
to that for
        $ \ell^\parallel_b $,
the continuum $\vk $-space analog;
        see Equation~(\ref{eq:corr-len-FT}).
However, in cases where there is no 2D energy one has
        $ \Ered( 0) = 0 $;
the choice
        $ X = \pi $
then gives the inappropriate result $ \ell_3 = 0 $.
Although mathematically correct this is clearly physically
misleading.  One can rectify this anomalous feature by considering
instead the limit as
        $ k_z \to 0 $
        \citep{MattGold82-rugged},
or using an $X$ value a little smaller than $\pi$ so that the
        $ \sin( k_3 X ) $
terms contribute;
       e.g., $ X = 9\pi / 10$.
%===============================================================

   \bigskip
        \section{Glossary}
                \label{sec:glossary}

Some definitions, more or less precise, of well-used/well-known
phrases associated with turbulence.
\begin{description}
 \item[toroidal]
        Polarized in the $ \vk \times \vB_0$ direction.
 \item[poloidal]
        Polarized in the $ \vk \times (\vk \times \vB_0) $ direction.

 \item[transverse]
        Perpendicular to $\vB_0$. Less restrictive than toroidal.

 \item[strong $B_0$]
        The fluctuations are energetically weak relative to the
        large-scale magnetic field:
         $ \delta v, \delta b \ll B_0 $.

 \item[Alfv\'enic]
        Used in several ways in the literature. For example\\
             (a) exactly like Alfv\'en waves (arguably a less common usage);\\
             (b) similar to or suggestive of Alfv\'en wave features
                 [near extremal $\sigma_c$,
                   approximate equipartition of kinetic and magnetic energy,
                   $ \delta \rho / \rho \ll 1 $].\\
        See Appendix~\ref{app:Alfvenic} for other usages.

 \item[2D modes]
   Those (Fourier) modes with $k_z = 0 $, so that they have no
   dependence on the parallel coordinate.

 \item[strict 2D]
    2D plus $ v_z = b_z = 0 $.  Hence polarized in the 2D plane.

 \item[quasi-2D]
   Modes with $k_z \approx 0 $ in the sense that
   (i) $ k_\perp \gg k_z $,
   and
   (ii) the associated linear Alfv\'en wave timescale
        $ \tauA = 1 / | \vk \cdot \vB_0 | $ is
    longer than the nonlinear timescale, $ \tauNL(k)$.

  \item[weak turbulence]
     At leading-order, fluctuations are waves, or quite wavelike.
     Nonlinear effects accumulate over many wave timescales.
     The fluctuations have a wave timescale that is fast compared to
     the nonlinear timescale:
                $ \tauW(\vk) \ll \tauNL( k) $.

  \item[strong turbulence]
     Nonlinear activity is strong, often dominant,
          with
                $ \tauW(\vk) \gtrsim \tauNL( k) $.
          Linear wave(like) activity might be comparable
          (energetically) to that of the nonlinear dynamics.

 \item[slab fluctuations]
        Those with their wavevectors strictly parallel to $\vB_0$.

 \item[slab turbulence]
    Used to mean a collection of slab fluctuations with
    different $\vk$s, all with $ \vk \times \vB_0 = 0 $.
    In fact, for incompressible systems the phrase is a misnomer:
    solenoidality of $\vv$ and $\vb$ means the nonlinear terms
    are exactly zero for \emph{any} set of slab fluctuations.
    Hence incompressible slab fluctuations obey a linear (Alfv\'en)
    wave dynamics,
    with no turbulence or spectral transfer involved.
\end{description}
%--------------------------------------------------------------

   \bibliographystyle{apj}
%   \bibliography{ag,hl,mp,qz}

  \providecommand{\SortNoop}[1]{} %.......Use as {\SortNoop{Aaa}}
  \providecommand{\sortnoop}[1]{} %..............................
  \newcommand{\stereo} {\emph{{S}{T}{E}{R}{E}{O}}} %.................
  \newcommand{\au} {{A}{U}\ } %..................................
  \newcommand{\MHD} {{M}{H}{D}\ } %..............................
  \newcommand{\mhd} {{M}{H}{D}\ } %...............................
  \newcommand{\RMHD} {{R}{M}{H}{D}\ } %...........................
  \newcommand{\rmhd} {{R}{M}{H}{D}\ } %...........................
  \newcommand{\wkb} {{W}{K}{B}\ } %..............................
  \newcommand{\alfven} {{A}lfv{\'e}n\ } %...........................
  \newcommand{\alfvenic} {{A}lfv{\'e}nic\ } %.........................
  \newcommand{\Alfven} {{A}lfv{\'e}n\ } %...........................
  \newcommand{\Alfvenic} {{A}lfv{\'e}nic\ }

%===============================================================

\end{document}